\documentclass[conference]{IEEEtran}
\IEEEoverridecommandlockouts
\usepackage{hyperref}
\usepackage{amsmath,amssymb,amsfonts, amsthm}
\usepackage{algorithm}
\usepackage{algpseudocode}
\usepackage{enumitem}
\usepackage{graphicx}
\usepackage{textcomp}
\usepackage{xcolor}
\def\BibTeX{{\rm B\kern-.05em{\sc i\kern-.025em b}\kern-.08em
    T\kern-.1667em\lower.7ex\hbox{E}\kern-.125emX}}
\newtheorem{theorem}{Theorem}
\newtheorem{lemma}[theorem]{Lemma}
\newtheorem{corollary}[theorem]{Corollary}
\newtheorem{assumption}{Assumption}
\newtheorem{remark}{Remark}
\newtheorem*{proofsketch}{Proof (Sketch)}

\newcommand{\E}{\mathbb{E}}
\newcommand{\PP}{\mathbb{P}}
\newcommand{\Pu}{P_{n,u,x}}
\newcommand{\R}{\mathbb{R}}
\newcommand{\cN}{\mathcal{N}}
\newcommand{\cM}{\mathcal{M}}
\newcommand{\cB}{\mathcal{B}}

\newcommand{\dd}{\,\mathrm{d}}
\newcommand{\rmin}{r_{\cM}}
\newcommand{\cbar}{\bar{c}}

\newcommand{\Binv}{B_{k,s}^{-1}}
\newcommand{\Var}{\operatorname{Var}}
\DeclareMathOperator*{\esssup}{ess\,sup}

\begin{document}

\title{On the Estimation of Chernoff Information}

\author{\IEEEauthorblockN{Kadircan Aksoy}
\IEEEauthorblockA{\textit{Institute for Space Research} \\
\textit{German Aerospace Center (DLR)}}
\IEEEauthorblockA{\textit{Chair of Communications and Information Theory} \\
\textit{Technische Universität Berlin}\\
Berlin, Germany \\
kadircan.aksoy@dlr.de}
\and
\IEEEauthorblockN{Peter Jung}
\IEEEauthorblockA{\textit{Institute for Space Research} \\
\textit{German Aerospace Center (DLR)}}
\IEEEauthorblockA{\textit{Chair of Communications and Information Theory} \\
\textit{Technische Universität Berlin}\\
Berlin, Germany \\
peter.jung@dlr.de}
}
\maketitle

\begin{abstract}
Chernoff information is a fundamental divergence measure characterizing the optimal error exponent in Bayesian binary hypothesis testing, with applications in information fusion, time-series analysis, and statistical learning theory. However, closed-form expressions exist only for simple parametric families, and nonparametric estimation remains difficult because the quantity is defined as an optimization of the unnormalized Rényi divergence over its order. We reformulate this optimization via a derivative condition, whose zero locates the optimal mixture parameter, and estimate the derivative directly using a $k$-nearest-neighbor method. We prove the $L_2$-consistency of the derivative estimator under mild regularity conditions on the densities and their domain. Coupled with a bisection procedure that locates the optimal parameter up to arbitrary precision, this yields an estimator for Chernoff information.
\end{abstract}


\section{Introduction}\label{sec:intro}

Binary hypothesis testing is one of the most fundamental problems in statistics. Given a sequence of random observations $X_1, \ldots, X_n \overset{\text{i.i.d.}} \sim X$, we wish to decide between two hypotheses $H_0: X \sim P$ vs. $H_1: X \sim Q$. Any decision rule $\delta_n : \mathcal{X}^n \to \{0,1\}$ incurs two types of error: the type-I error $\alpha_n(\delta_n) = P^{\otimes n}(\delta_n = 1)$; the probability of falsely rejecting $H_0$, and the type-II error $\beta_n(\delta_n) = Q^{\otimes n}(\delta_n = 0)$; the probability of failing to detect $H_1$. There is an inherent tradeoff between the two, no decision rule can drive both to zero simultaneously for finite $n$, and the central question is how fast they can be made to decay as $n \to \infty$. \\

In the asymmetric setting, the Neyman--Pearson lemma \cite{np} establishes that fixing $\alpha_n \leq \varepsilon$ and minimizing $\beta_n$ is achieved by a likelihood ratio threshold test, with the optimal $\beta_n$ decaying at an exponential rate $e^{-n D_{KL}(P \| Q)}$ \cite{stein}, where
\begin{align}
    D_{KL}(P \| Q) := \int_{\mathcal X} p(x) \log \frac{p(x)}{q(x)} \, d\mu(x)
\end{align}
is the Kullback--Leibler divergence between $P$ and $Q$ \cite{kl}. In the symmetric Bayesian setting with priors $\pi = (\pi_0, \pi_1)$, we instead minimize the total probability of error:
\begin{align}
    P_e^{(n)*}(\pi) = \inf_{\delta_n}\left[\pi_0\, \alpha_n(\delta_n) + \pi_1\, \beta_n(\delta_n)\right].
\end{align}

A classical result due to Chernoff \cite{chernoff} establishes that $P_e^{(n)*}(\pi)$ also decays exponentially in $n$, but with an exponent that is independent of the prior $\pi$. In fact, for any $\pi_0, \pi_1 \in (0,1)$ with $\pi_0 + \pi_1 = 1$ we have
\begin{align}
    \lim_{n \to \infty} -\frac{1}{n} \log P_e^{(n)*}(\pi) = C(P, Q),
\end{align}
where $C(P, Q)$, called the \textbf{Chernoff information}, is defined as:
\begin{align}
    C(P, Q) &:= \sup_{s \in [0,1]} -\log  C_s(P, Q), \ \text{s.t.}\\
    C_s(P, Q) &:= \int_{\mathcal{X}} p(x)^s q(x)^{1-s} \, d\mu(x),
\end{align}
with $C_s(P, Q):[0,1] \to (0,1]$ \footnote{The range can be obtained via a simple application of Hölder's inequality \cite{durrett}), also see \cite{nielsen}.} denoting the \textbf{Chernoff Functional} of order $s$, assuming $P$ and $Q$ are commonly dominated by a measure $\mu$ on $\mathcal{X}$. \\

The Chernoff functional is directly related to the R\'enyi divergence \cite{renyi} of order $s$,
\begin{align}
    R_s(P \| Q) := \frac{1}{s-1} \log C_s(P,Q),
\end{align}
through
\begin{align}
    -\log C_s(P,Q) = (1-s) R_s(P \| Q),
\end{align}
so that
\begin{align}
    C(P,Q) = \sup_{s \in [0,1]} (1-s) R_s(P \| Q),
\end{align}
meaning that Chernoff information is the maximum of the R\'enyi divergence with its $(1-s)^{-1}$ normalization removed. \\

Since $\log C_s(P, Q)$ is strictly convex whenever $P \neq Q$,\footnote{The second derivative $\frac{d^2}{ds^2}\log C_s(P,Q) = \operatorname{Var}_{P_s}[\log(p/q)]$ is strictly positive unless $\log(p/q)$ is $\mu$-a.e. constant.} with $\log C_0(P, Q) = \log C_1(P, Q) = 0$, the supremum is attained at an interior point $s^* \in (0,1)$ with $C(P,Q) > 0$. Therefore the optimal $s^*$ is the \textit{unique} solution to
\begin{align} \label{Cs-diff}
    \frac{d}{ds} \log C_s(P, Q)\bigg|_{s=s^*} = \mathbb{E}_{P_{s^*}}\!\left[\log\frac{p(X)}{q(X)}\right] = 0,
\end{align}
where 
\begin{align}
    p_s(x) = \frac{p(x)^s q(x)^{1-s}}{\int_{\mathcal X} p(x)^s q(x)^{1-s} \, d\mu(x)}
\end{align}
is the tilted density at parameter $s$. Interestingly, at the optimum $s^*$, $C(P,Q)$ attains the following relation with KL divergence: 
\begin{align}
    C(P,Q) = D_{KL} (P_{s^*} \| P) = D_{KL} (P_{s^*} \| Q)
\end{align}

\section{Related Work}

\subsection{$k$-Nearest-Neighbor Density Estimation}

$k$-nearest-neighbor ($k$-NN)-based density estimators \cite{fukunaga} exploit a simple geometric fact: in a region of $\mathbb{R}^d$ where the density $p$ is approximately locally constant, the volume of the ball containing the $k$ nearest neighbors of a point $x$ among $n$ samples is approximately $k/(np(x))$. Inverting this relationship yields a pointwise, non-parametric density estimate, which is convergent when $k$ grows sub-linearly with $n$ \cite{loftsgaarden}, \cite{gao}. \footnote{Note for finite $k$, pointwise estimates can be biased and must be corrected depending on the quantity of interest where the density estimate is used. }\\

Formally, let $X_1, \ldots, X_n \overset{\text{i.i.d.}}{\sim} P$ and let $\rho_{k,n}(x)$ denote the Euclidean distance from $x$ to its $k$-th nearest neighbor in $\{X_1, \ldots, X_n\}$. The $k$-NN density estimate of $p$ at $x$ is
\begin{align}
    \widehat{p}_{k,n}(x) = \frac{k}{n \cdot \cbar \cdot \rho_{k,n}(x)^d},
\end{align}
where $\cbar = \pi^{d/2} / \Gamma(d/2 + 1)$ is the volume of the unit ball in $\mathbb{R}^d$ and $\Gamma(z) = \int_0^\infty t^{z-1} e^{-t} \, dt$ denotes the gamma function. Intuitively, the ball of radius $\rho_{k,n}(x)$ centered at $x$ contains exactly $k$ sample points, so the empirical mass in that ball is $k/n$, giving the density estimate above.

\subsection{kNN Estimate of the Chernoff Functional}
Recall that 
\begin{align}
C_s(P,Q) := \int_{\mathcal X} p(x)^s q(x)^{1-s}\,d\mu(x) = \mathbb{E}_{X\sim P}\left[\left(\frac{q(X)}{p(X)}\right)^{1-s}\right]
\end{align}

Given i.i.d samples $X^n = \{X_1, X_2 ... X_n\}$, $X_i \overset{\text{i.i.d.}} \sim P$ and $Y^m = \{Y_1, Y_2 ... Y_m\} \overset{\text{i.i.d.}} \sim Q$, let \(\rho_k(i)\) be the distance from \(X_i\) to its $k$-th nearest neighbor among \(\{X_j\}_{j\neq i}\), and let \(\nu_k(i)\) be the distance from \(X_i\) to its $k$-th nearest neighbor among \(\{Y_j\}_{j=1}^m\). Using the $k$-NN density-ratio approximation described above we can write
\begin{align}
\frac{q(X_i)}{p(X_i)} \approx \frac{(n-1)\rho_k(i)^d}{m\nu_k(i)^d} := r_i,
\end{align}

\cite{PS} uses $r_i$ to estimate $C_s(P,Q)$ by averaging over the $P$-samples. The resulting ($L_2$-consistent) estimator has the form 
\begin{align} \label{poczos-estimator}
\widehat C_s(X^n,Y^m) =& \frac{1}{n}\sum_{i=1}^n B_{k,s} r_i^{1-s}
\end{align}
where
\begin{align}
B_{k,s} =& \frac{\Gamma(k)^2}{\Gamma(k-s+1)\Gamma(k+s-1)}.
\end{align}

The multiplicative factor $B_{k,s}$ is the finite-$k$ correction term arising from the limiting Erlang law of $k$-NN volume terms $r_i$ (see proofs in Appendix, particularly Lemma~\ref{lem:cdf} and~\ref{lem:erlang} for a derivation).

\section{Proposed Estimator}
We wish to construct an estimator $\widehat{C}(X^n, Y^m)$ for the true Chernoff information $C(P,Q)$ such that it is $L_2$-consistent:
\begin{align}
    \lim_{n,m \to \infty} &\mathbb{E}[\widehat{C}(X^n, Y^m) ] = C(P,Q) \, \\
    \lim_{n,m \to \infty} &\mathbb{E}[(\widehat{C}(X^n, Y^m) - C(P,Q))^2] = 0.
\end{align}

\cite{nielsen} has shown that one can approximate $s^*$ up to arbitrary precision $|s^* - \widehat{s}| < \varepsilon$ in $\mathcal{O}(\log\frac{1}{\varepsilon})$ time by exploiting the \textit{information geometry} (see \cite{amari} for a detailed reference on the subject) with a simple bisection algorithm. Yet this depends crucially on the step of comparing two KL terms: $D_{KL} (P_{\widehat{s}} \| P)$ vs $D_{KL} (P_{\widehat{s}} \| Q)$ for the current guess of the mixing parameter $\widehat{s}$. In general, if the densities $p$ and $q$ are available, one can compute this analytically (given the integral is tractable) but from merely samples, it is non-trivial to sample from the mixed distribution $P_{\widehat{s}}$ and re-utilize a KL divergence estimator, e.g from \cite{verdu}. \\

\cite{cine} overcomes this by constructing a neural estimator for $C_s(P,Q)$, similar to MINE for mutual information \cite{mine} and solving the convex optimization over $s$ directly. The consistency of this estimator however, crucially depends on an assumption that the density ratio estimate $r_i$ is itself consistent. Neural estimators of this kind such as MINE, are known to suffer from high variance and bias that grows with the true divergence \cite{minelimits}. \cite{cine} likewise reports degradation and sensitivity to hyperparameters. \\

Instead, we rewrite the bisection condition from \cite{nielsen} as the derivative condition from equation \eqref{Cs-diff}. Note that:
\begin{align}
    & D_{KL} (P_{\widehat{s}} \| P) > D_{KL} (P_{\widehat{s}} \| Q) \\
    \iff & \mathbb{E}_{P_{\widehat{s}}}\!\left[\log\frac{p(X)}{q(X)}\right] < 0  \\
    \iff & D_s(P,Q) := \frac{d}{ds} \log C_s(P,Q) \bigg|_{s=\widehat{s}} < 0
\end{align}

Therefore, one can instead check the sign of the derivative in the bisection step. A natural estimate of this derivative is via a finite difference
\begin{align}
\Delta_h(s) := \frac{\widehat C_{s+h} - \widehat C_{s-h}}{2h},
\qquad h>0.
\end{align}
However, this approach is not reliable due to the unknown finite-sample bias of $\widehat C_s$. To make this precise, write the estimator as
\begin{align}
\widehat C_s = C_s + b_n(s),
\end{align}
where $b_n(s)$ denotes an unknown bias term. Then
\begin{align}
\Delta_h(s)
&= \frac{C_{s+h} - C_{s-h}}{2h} + \frac{b_n(s+h) - b_n(s-h)}{2h}.
\end{align}
Assuming $C_s$ is twice continuously differentiable, a Taylor expansion yields
\begin{align}
\frac{C_{s+h} - C_{s-h}}{2h} = C_s'(s) + \mathcal{O}(h^2),
\end{align}
so that
\begin{align}
\Delta_h(s) = C_s'(s) + \mathcal{O}(h^2) + \frac{b_n(s+h) - b_n(s-h)}{2h}.
\end{align}

Hence the sign of $\Delta_h(s)$ coincides with the sign of the true derivative $C_s'(s)$ only if
\begin{align}
\left|C_s'(s)\right|
>
\left|\mathcal{O}(h^2)\right|
+
\left|
\frac{b_n(s+h) - b_n(s-h)}{2h}
\right|.
\end{align}

The difficulty arises because the bias term has an unknown relation to $s$. In particular, even if $b_n(s)$ is small in magnitude, the difference quotient
\begin{align}
\frac{b_n(s+h) - b_n(s-h)}{2h}
\end{align}
may become arbitrarily large as $h$ shrinks, unless $b_n$ is sufficiently regular in $s$. \\

Indeed, near the optimum $s^*$, we have $C_s'(s^*) \approx 0$, so the above condition fails more dramatically in a neighborhood of $s^*$, as the sign of $\Delta_h(s)$ is increasingly determined by the bias-induced term. For high precision applications this may be fatal. Moreover, decreasing $h$ amplifies the bias contribution through the factor $1/h$, while increasing $h$ increases the deterministic approximation error $\mathcal{O}(h^2)$. \\

To combat this, we develop an estimator $\widehat{D_s}(X^n, Y^m)$ directly for the derivative, by differentiating the estimator from \cite{PS}:
\begin{align}
    \widehat{C_s}(X^n, Y^m) = \frac{1}{n} \sum B_{k,s} r_i^{1-s}
\end{align}

We begin by simply noting that 
\begin{align}
    \frac{d}{ds} \log \widehat{C_s}(X^n, Y^m) = \frac{\frac{d}{ds} \widehat{C_s}(X^n, Y^m)}{\widehat{C_s}(X^n, Y^m)}
\end{align}

Then we have for the numerator:
\begin{align}
   \frac{d}{ds} \widehat{C_s}(X^n, Y^m) &= \frac{1}{n}\sum_{i=1}^n \partial_s B_{k,s} r_i^{1-s} - B_{k,s} \log r_i r_i^{1-s}
\end{align}

Taking the $\log$ of $B_{k,s}$
\begin{align}
   \log B_{k,s} &= 2\log \Gamma(k) - \log \Gamma(k + 1 - s) \\& \quad - \log \Gamma(k - 1 + s) \\
   \implies \partial_s \log B_{k,s} &= \psi(k + 1 - s) - \psi(k - 1 + s) \\
   \implies  \partial_s B_{k,s} &= B_{k,s} [\psi(k + 1 - s) - \psi(k - 1 + s)]
\end{align}

Bringing the two terms together and noting that the $B_{k,s}$ factors cancel, we obtain the derivative estimator: 
\begin{align}\label{eqn:Dshat}
    \widehat{D_s}(X^n, Y^m) &= G_{k,s} - \frac{\frac{1}{n}\sum_{i=1}^n r_i^\gamma \log r_i}{\frac{1}{n}\sum_{i=1}^n r_i^\gamma} \ , \text{where} \\
    G_{k,s} & := \psi(k + \gamma) - \psi(k - \gamma) \ \text{and } \gamma := 1-s.
\end{align}
Combining $\widehat{D_s}$ with the bisection algorithm from \cite{nielsen}, we get the procedure in Algorithm $1$. \\

\begin{algorithm}
\caption{Sample based estimator for Chernoff information}
\begin{algorithmic}[1]
\Require Two sample sets $X^n = \{X_1, X_2 ... X_n\}$, $X_i \overset{\text{i.i.d.}} \sim P$ and $Y^m = \{Y_1, Y_2 ... Y_m\}, Y_j \overset{\text{i.i.d.}} \sim Q$, precision parameter $\varepsilon > 0$, integer $k \ge 5$
\State $s_m \leftarrow 0$,\quad $s_M \leftarrow 1$
\State $r_i \leftarrow \frac{(n-1)\rho_k(i)^d}{m\nu_k(i)^d}$ with $\rho_k(i), \nu_k(i)$ the distance of point $X_i$ to its k-th neighbour in $X^n \setminus X_i$ and $Y^m$, respectively.
\While{$|s_M - s_m| > \varepsilon$}
    \State $\hat s \leftarrow (s_m + s_M) / 2, \; \gamma \leftarrow 1 - \hat s$
    \State $G_{k,s} \leftarrow \psi(k + \gamma) - \psi(k - \gamma)$
    \State $\widehat D_s \leftarrow G_{k,s} - \frac{\frac{1}{n}\sum_{i=1}^n r_i^\gamma \log r_i}{\frac{1}{n}\sum_{i=1}^n r_i^\gamma}$, k-NN derivative estimate at $\hat s$
    \If{$\widehat D_s > 0$}
        \State $s_M \leftarrow \hat s$
    \Else
        \State $s_m \leftarrow \hat s$
    \EndIf
\EndWhile
\State $B_{k, s} \leftarrow \frac{\Gamma(k)^2}{\Gamma(k+\gamma)\Gamma(k-\gamma)}$
\State $C \leftarrow -\log (B_{k,s} \frac{1}{n}\sum_{i=1}^n r_i^\gamma)$, k-NN estimate of Chernoff information
\State \Return $C$
\end{algorithmic}
\end{algorithm}

A priori, it is not the case that consistency of $\widehat C_s$ implies the consistency of its derivative. However, for our particular $\widehat D_s$, this turns out to be the case, which we now show.

\section{Asymptotic Consistency}

Throughout the rest of the text, let $\cM \subseteq \R^d$ be a domain such that $\text{supp}(P) = \text{supp}(Q) = \cM$, with densities $p = dP/d\mu$ and $q = dQ/d\mu$ against the Lebesgue measure $\mu$ over $\cM$. We make the following standing assumptions:

\begin{assumption}[Bounded domain]\label{ass:bdd}
Let $\cM$ be bounded with diameter $D := \sup_{x,y\in\cM}\|x-y\|$.
\end{assumption}

\begin{assumption}[Bounded densities]\label{ass:dens}
Let $C_p,C_q,c_p,c_q \in \R_{>0}$ be constants s.t, for almost all $x\in\cM$,
\begin{align}
  c_p < p(x) < C_p, \qquad c_q < q(x) < C_q.
\end{align}
\end{assumption}

\begin{assumption}[Volume regularity of the domain]\label{ass:vol} Let $\rmin >0$ such that
\begin{align}
  \inf_{0<\delta\leq1}\,\inf_{x\in\cM}\,
  \frac{V\left(\cB(x,\delta)\cap\cM\right)}{V\left(\cB(x,\delta)\right)} = \rmin
\end{align}
i.e, over a shrinking sequence of balls, the domain $\cM$ can not shrink arbitrarily quickly.
\end{assumption}

\begin{remark}\label{rem:assumptions}
The bounded-domain and lower-bound assumptions are needed to keep the $\log r_i$ terms finite. \cite{verdu} instead assumes $\mu$-regularity, a set of $\mu^{th}$-moment conditions on $\log p(x)$ and $\log\|x-y\|$ that keep $\log r_i$ finite in expectation. However \cite{PS} points out that the proof in \cite{verdu} under this condition is flawed, as it applies the reverse Fatou lemma (see \cite{lieb}) under conditions where it does not hold. As a result, adapting $\mu$-regularity in place of Assumptions~\ref{ass:bdd} and~\ref{ass:dens} could relax both, but we retain them for proof simplicity.
\end{remark}

We now state moment bounds on the $k$-NN volume terms, which will be needed for asymptotic consistency proofs.

\begin{lemma} [Moments]\label{lem:rmom}
$\forall |\beta|<k$, we have
\begin{align}
  \sup_{n,m,i}\,\E\left[\,r_i^{\,\beta}\,\right] < \infty .
\end{align}
\end{lemma}

\begin{lemma}[Log-weighted moments]\label{lem:logmom}
$\forall |\beta|<k, \theta\ge 0$, we have
\begin{align}
  \sup_{n,m,i}\,\E\left[\,r_i^{\,\beta}\,\bigl|\log r_i\bigr|^{\theta}\,\right]
  < \infty .
\end{align}
\end{lemma}

The proof for both can be found in Appendix~\ref{app:lemmaproofs}. The key idea is to use the lower and upper bounds on the densities from Assumption~\ref{ass:dens} to bound the probability of $k$-NN ball volumes growing or shrinking arbitrarily fast. Combined with Assumption~\ref{ass:bdd}, which prevents the samples from being arbitrarily far apart, this yields the moment bounds.\\

Note that these lemmas introduce a constraint on the minimum $k$ that can be used. In particular; for retrieving $L_p$-consistency, we need to have $k>2p$. This fact becomes clear during the consistency proofs (see Appendix~\ref{app:thmproofs}). Therefore in practice, unless stated otherwise, we take $k = 5$ to ensure $L_2$-consistency and all theorem/lemmas are stated with $k \ge 5$. 

\subsection{$L_1$-Consistency}\label{sec:L1}

\begin{theorem}\label{thm:L1}
    For $k \ge 5$, the derivative estimator $\widehat D_s(X^n, Y^m)$ is $L_1$-consistent:
\begin{align}
    \lim_{n,m \to \infty} \mathbb{E}_{P^{\otimes n}, Q^{\otimes m}} \bigl[\,\bigl|\widehat D_s(X^n, Y^m) - D_s(P,Q)\bigr|\,\bigr] = 0.
\end{align}
\end{theorem}

\begin{proofsketch}
Fix $P,Q$ and $s$, and define
\begin{align*}
    a_s &= \Binv C_s(P,Q), \quad b_s = \Binv C_s(P,Q)\big( G_{k,s} - D_s(P,Q) \big) \\
    A_n &= \frac{1}{n} \sum_{i=1}^n r_i^{\,\gamma} \quad B_n = \frac{1}{n} \sum_{i=1}^n r_i^{\,\gamma} \log r_i.
\end{align*}
with $\widehat D_s = G_{k,s} - B_n/A_n$. We first analyze $A_n$ and $B_n$ separately to show $\mathbb E[A_n] \to a_s$ and $\mathbb E[B_n] \to b_s$. Then we establish the convergence of the ratio $\mathbb E[B_n/A_n] \to b_s/a_s$. \\

Since $A_n$ is exactly the estimator for $C_s(P,Q)$ from \cite{PS}, its consistency follows verbatim. Centrally, the pointwise kNN volume terms $r_i$ converge in distribution to Erlang laws $\mathrm{Erlang}(k, \cbar p(x_i))$ by a counting argument. Using the closed-form moments of these laws yields $\mathbb E[A_n] \to a_s$.\\

For $B_n$, we proceed similarly and differentiate the corresponding moment expressions. Justification for interchanging differentiation and expectation follows from dominated convergence, using an integrable envelope for the log-weighted terms. This yields $\mathbb E[B_n] \to b_s$.\\

To establish the convergence of the ratio, we show the following:
\begin{enumerate}
    \item Both terms have zero asymptotic variance: $\mathrm{Var}(A_n) \to 0$ and $\mathrm{Var}(B_n) \to 0$.
    \item By Chebyshev’s inequality \cite{durrett},
    \begin{align*}
        \mathbb P(|A_n - \mathbb E[A_n]| \ge \varepsilon) \leq \frac{\mathrm{Var}(A_n)}{\varepsilon^2} \to 0, \quad \forall \varepsilon > 0,
    \end{align*}
    hence $A_n \xrightarrow{\mathbb P} a_s$ and $B_n \xrightarrow{\mathbb P} b_s$ (since $\E[A_n] \to a_s$).
    \item For $a_s \neq 0$, the Continuous Mapping theorem \cite{dervaart} implies $\frac{B_n}{A_n} \xrightarrow{\mathbb P} \frac{b_s}{a_s}.$
    \item Finally, the ratio family $\{B_n / A_n\}_{n \geq 1}$ is uniformly integrable and by Vitali’s convergence theorem \cite{vitali}:
    \begin{align*}
        \mathbb E\!\left[\Bigl|\frac{B_n}{A_n} - \frac{b_s}{a_s}\Bigr|\right] \to 0.
    \end{align*}
\end{enumerate}

Plugging $a_s$ and $b_s$ back, we obtain
\begin{align}
    \lim_{n,m \to \infty} \mathbb{E} \bigg[ G_{k,s} - \frac{B_n}{A_n} \bigg] = D_s(P,Q),
\end{align}
after adding the correction term $G_{k,s}$.
\end{proofsketch}

\subsection{$L_2$-Consistency}
\begin{theorem}\label{thm:L2}
For $k \ge 5$, the derivative estimator $\widehat D_s(X^n, Y^m)$ is $L_2$-consistent:
\begin{align*}
    \lim_{n,m\to \infty} \mathbb E_{P^\otimes n, Q^\otimes m}[(\widehat D_s(X^n, Y^m) - D_s(P,Q))^2] = 0
\end{align*}
\end{theorem}

\begin{proofsketch}

We have already most of the ingredients needed to show this statement. Similar to Theorem~\ref{thm:L1}, we first show the $L_2$ convergence of $A_n$ and $B_n$ to then argue for the convergence of their ratio. \\

Let again $\widehat D_s = G_{k,s} - \frac{B_n}{A_n}$ and recall that $A_n \xrightarrow{\mathbb P} a_s$, $B_n \xrightarrow{\mathbb P} b_s$ with $a_s = \Binv C_s(P,Q)$ and $b_s = \Binv C_s(P,Q)\bigl(G_{k,s} - D_s(P,Q)\bigr)$. \\

Furthermore we have $\mathrm{Var}(A_n)\to 0$ and $\mathrm{Var}(B_n)\to 0$ together with convergence of the means; from these we obtain $A_n \xrightarrow{L_2} a_s$ and $B_n \xrightarrow{L_2} b_s$. \\

To apply Vitali's theorem again, we show the squared family $\left\{ \left(\frac{B_n}{A_n}\right)^2 \right\}_{n\geq 1}$ is also uniformly integrable where the proof goes analogously to the $L_1$ lemma, which together with convergence in probability implies $\frac{B_n}{A_n} \to \frac{b_s}{a_s}$ in $L_2$. Therefore we have
\begin{align}
&\lim_{n,m \to \infty}
\mathbb E\!\left[(\widehat D_s - D_s)^2\right]
=
\lim_{n,m \to \infty}
\mathbb E\!\left[\left(G_{k,s} - \frac{B_n}{A_n} - D_s\right)^2\right] \\
=&
\lim_{n,m \to \infty}
\mathbb E\!\left[\left(\frac{B_n}{A_n} - \frac{b_s}{a_s}\right)^2\right]
= 0.
\end{align}
\end{proofsketch}

\subsection{Consistency of the Bisection}
Theorems~\ref{thm:L1} and~\ref{thm:L2} establish the asymptotic consistency of the derivative estimator $\widehat D_s$ for any fixed $s$. This does not by itself imply that the bisection algorithm, driven by $\widehat D_s$, locates the correct root $s^*$. We conclude this section by showing that the bisection via $\widehat D_s$ also approximates $s^*$ to arbitrary precision.
    
\begin{theorem}[Consistency of the bisection]\label{thm:bisection_proof}
Let $s^* \in (0,1)$ be the unique root of $D_s$ and $\hat s_{n,m} \in (0,1)$, denote the bisection estimate with $n,m$ samples and precision parameter $\varepsilon > 0$. Then $\forall \eta > 0$ with $s^* \pm \eta \in (0,1)$,
\begin{align}
    \lim_{n,m \to \infty} \mathbb P(|\hat s_{n,m} - s^*| > \eta + \varepsilon) = 0.
\end{align}
\end{theorem}

\begin{proof}

    Fix $\eta > 0$ s.t $s^* \pm \eta \in (0,1)$ and define
    \begin{align}
        \Delta(\eta) := \min\{\,|D_{s^*-\eta}|,\ |D_{s^*+\eta}|\,\} > 0 .
    \end{align}
    By monotonicity of $D_s$ (established in Section \ref{sec:intro}), every $s$ with $|s - s^*| \ge \eta$ satisfies $|D_s| \ge \Delta(\eta)$. The trick is to see that only a sign flip by the estimator results in large error. Concretely, if $\operatorname{sign}(\widehat D_s) \ne \operatorname{sign}(D_s)$, then $\widehat D_s$ and $D_s$ lie on opposite sides of $0$. We get
    \begin{align}
        \operatorname{sign}(\widehat D_s) \ne \operatorname{sign}(D_s) \implies |\widehat D_s - D_s| \ge |D_s| \ge \Delta(\eta).
    \end{align}
    Taking the probabilities of each event and applying Markov's inequality along with $L_1$-consistency (Theorem \ref{thm:L1}) gives, $ \forall s$ s.t $|s - s^*| \ge \eta$:
    \begin{align}
        \mathbb{P}\big(\operatorname{sign}(\widehat D_s) \ne \operatorname{sign}(D_s)\big)
        &\le \mathbb{P}\big(|\widehat D_s - D_s| \ge \Delta(\eta)\big) \\
        &\le \frac{\E[|\widehat D_s - D_s|]}{\Delta(\eta)} \to 0,
    \end{align}
    
    Note that this holds at any fixed $s$ with $|s - s^*| \ge \eta$. Since the bisection runs for $T = O(\log\frac1\varepsilon)$ steps, there are at most $2^T$ distinct points the algorithm can visit, and which of them it visits depends on the samples. A union bound over all of them gives
    \begin{align}
        \mathbb{P}(|\hat s_{n,m} - s^*| > \eta + \varepsilon)
        \le 2^T\ \frac{\max_s \E[|\widehat D_s - D_s|]}{\Delta(\eta)} \to 0,
    \end{align}
    as $T$ is finite and the maximum is taken over the finitely many points the algorithm can visit. If no sign flip occurs at any $s$ with $|s-s^*| \ge \eta$, the interval $[s_m, s_M]$ always contains a point within $\eta$ of $s^*$, and it has width at most $\varepsilon$ on termination, giving $|\hat s_{n,m} - s^*| \le \eta + \varepsilon$.

\end{proof}

\section{Experimental Results}
We conduct experiments on simple datasets to verify the convergence of the estimator for feasible sample sizes and distributions. Appendix~\ref{app:exp} contains more plots for different parameter settings for the families described here.

\subsection{Symmetric Truncated Gaussians}

First, we study a symmetric pair of truncated Gaussians on $\cM = [0,1]$, as a closed-form sanity check. For mean $\mu$ and deviation $\sigma > 0$, let $\cN_{[0,1]}(\mu,\sigma)$ denote the Gaussian truncated to $[0,1]$. Note that it has density
\begin{align}
    p_{\mu,\sigma}(x) = \frac{1}{\sigma\, Z(\mu,\sigma)}\,
    \phi\!\left(\frac{x-\mu}{\sigma}\right) \\
    Z(\mu,\sigma) := \Phi\!\left(\frac{1-\mu}{\sigma}\right) - \Phi\!\left(\frac{-\mu}{\sigma}\right),
\end{align}
where $\phi, \Phi$ are the standard normal density and CDF. Pick a $\delta \in (0, 1/2)$ and set
\begin{align}
    P \sim \cN_{[0,1]}\!\left(\tfrac12 - \delta,\, \sigma\right), \qquad
    Q \sim \cN_{[0,1]}\!\left(\tfrac12 + \delta,\, \sigma\right),
\end{align}

It is easy to check that Assumption~\ref{ass:bdd} and~\ref{ass:vol} hold with $D = 1$ and $r_\cM = 1/2$. Moreover, both densities are continuous and strictly positive on the compact set $[0,1]$, hence bounded above and away from zero (Assumption~\ref{ass:dens}). \\

Since the densities are symmetric, i.e $q(x) = p(1-x), \forall x \in [0,1]$, we have that $C_s$ is symmetric about $s = 1/2$. As $C_s$ is also convex with $C_0 = C_1 = 1$, we immediately get $s^* = 1/2$. Therefore we can express $C(P, Q)$ as: 
\begin{align}
    C(P,Q) &= -\log C_{1/2}(P,Q) \\
    &= -\log \int_0^1 \sqrt{p(x)\,p(1-x)}\,dx \notag \\
    &= -\log \int_0^1 \frac{e^{-\delta^2/2\sigma^2}}{\sigma\sqrt{2\pi}\,Z(\mu,\sigma)}\,
       e^{-(x-1/2)^2/2\sigma^2}\,dx \notag \\
    &= \frac{\delta^2}{2\sigma^2}
       + \log Z(\tfrac12 - \delta,\sigma) - \log Z(\tfrac12,\sigma).
\end{align}
\begin{figure}
    \centering
    \includegraphics[width=\linewidth]{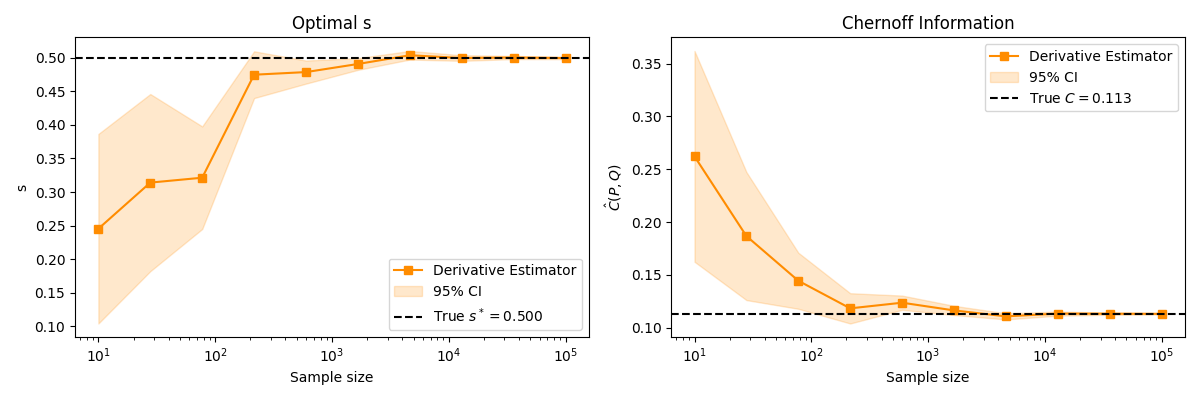}
    \caption{Estimator performance vs. sample size for $P \sim \cN_{[0,1]}\!\left(0.4,\, 0.2\right)$ and $Q \sim \cN_{[0,1]}\!\left(0.6,\, 0.2\right)$. The bisection is performed for 10 iterations for each sample size. We observe convergence around $10^3 - 10^4$ samples. 95\% confidence intervals (over 50 trials) are also reported.}
    \label{fig:symmetric_truncnorm}
\end{figure}

Figure \ref{fig:symmetric_truncnorm} shows the convergence rate of the estimator for $\delta = 0.1$ and $\sigma=0.2$. Indeed, for $\sim10^4$ samples, we observe $\widehat{D_s}, \hat{s^*}$ and $ \widehat C(P,Q)$ are all within $10^{-4}$ of their ground truth values.

\subsection{Truncated Exponentials}
As a canonical family, we study truncated exponentials. We fix the domain again to be the unit interval $\cM = [0,1]$ and let $Exp_{[0,1]}(\lambda)$ denote the exponential law with parameter $\lambda > 0$ and density (on $[0,1]$):
\begin{align}
    p_\lambda(x) = \frac{\lambda e^{-\lambda x}}{Z(\lambda)}, \quad
    Z(\lambda) = \int_0^1 \lambda e^{-\lambda x}\,dx = 1 - e^{-\lambda}
\end{align}

Take $P \sim Exp_{[0,1]}(\lambda_1)$ and $Q \sim Exp_{[0,1]}(\lambda_2)$ for $\lambda_1,\lambda_2 > 0$. Then both $C_s$ and $D_s$ have closed forms: 
\begin{align}
    C_s(P,Q) &= \int_0^1 p(x)^s q(x)^{1-s}\,dx \\
    &= \frac{\lambda_1^s \lambda_2^{1-s}}{Z(\lambda_1)^s Z(\lambda_2)^{1-s}}\frac{1 - e^{-\lambda(s)}}{\lambda(s)}.
\end{align}
with $\lambda(s) := s\lambda_1 + (1-s)\lambda_2$, the geometric mixture of the parameters. Taking logarithms and differentiating in $s$, with $\lambda'(s) = \lambda_1 - \lambda_2$,
\begin{align}
    D_s(P,Q) = \log\frac{\lambda_1 Z(\lambda_2)}{\lambda_2 Z(\lambda_1)}
      + (\lambda_1 - \lambda_2)\left[
      \frac{e^{-\lambda(s)}}{1 - e^{-\lambda(s)}} - \frac{1}{\lambda(s)}\right].
\end{align}

$D_s = 0$ is a transcendental equation, so the analytic form of $s^*$ is not known, but it can be approximated to machine precision by a bisection procedure similar to our Algorithm 1. Note that this is on the \textit{exact} value of $D_s$, unlike our sample based estimator $\widehat{D_s}$.\\

\begin{figure}
    \centering
    \includegraphics[width=\linewidth]{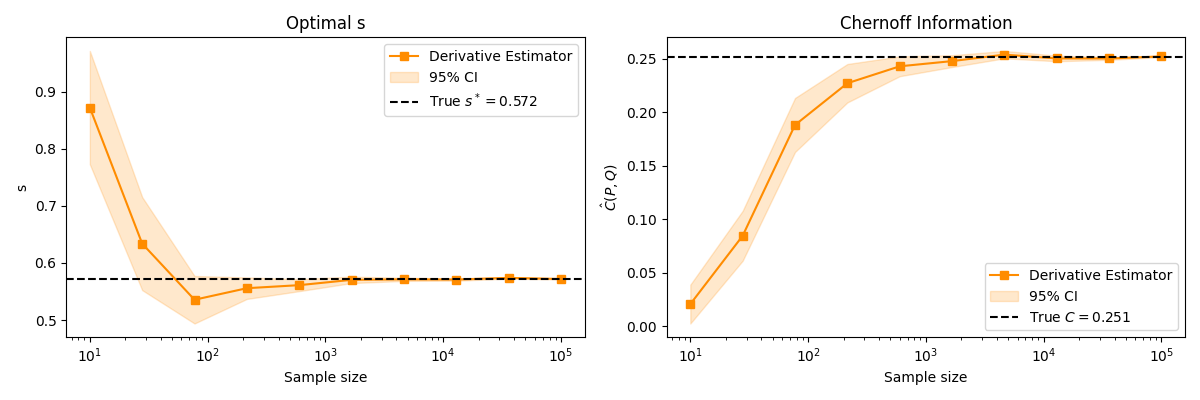}
    \caption{Estimator performance vs. sample size for $P \sim Exp_{[0,1]}(1.0)$ and $Q \sim Exp_{[0,1]}(8.0)$, restricted on the unit interval $\cM = [0,1]$. The bisection is performed for 10 iterations for each sample size. Convergence is around $10^3 -10^4$. Shaded areas are 95\% confidence intervals over 50 trials.}
    \label{fig:truncexp}
\end{figure}

\begin{figure}
    \centering
    \includegraphics[width=\linewidth]{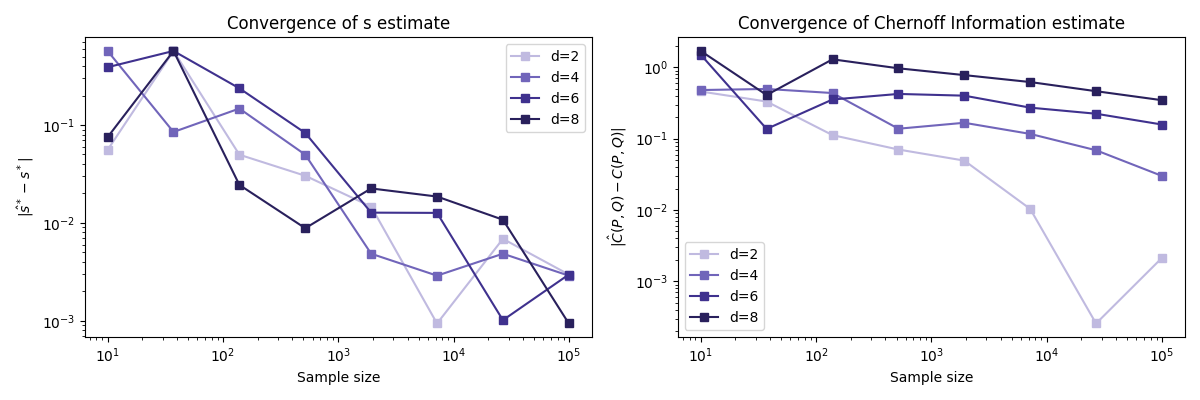}
    \caption{Convergence rate vs dimensionality of the distributions. x-axis is the number of samples while y is the error from ground truth for both $s^*$ and $C(P,Q)$. Each curve is for a pair of distributions on $\cM^d = [0,1]^d$ with $P \sim Exp_{[0,1]}([1.0]^d)$ and $Q \sim Exp_{[0,1]}([8.0]^d)$. Interestingly, as dimensions increase the convergence rate for the $s$ stays the same while the estimated $\widehat{C}_s$ starts to decline in quality.}
    \label{fig:truncexp_dims}
\end{figure}

Figure~\ref{fig:truncexp} shows results for $\lambda_1 = 1.0$ and $\lambda_2 = 8.0$, with convergence rates similar to the truncated Gaussian case. Figure~\ref{fig:truncexp_dims} extends to higher-dimensional densities on $\cM^d = [0,1]^d$ and shows the effect of dimension on the number of samples required for convergence, suggesting a sample complexity exponential in the dimension $d$ (see the Conclusion for further discussion), as is expected to be inherited from the $k$-NN density-ratio estimator. Interestingly, the $s^*$ search converges at roughly the same rate across the tested dimensions. Since the bisection requires only the sign of $\widehat{D_s}$, the root location may be more robust to the dimension-driven error growth that affects the divergence estimate itself.

\subsection{MNIST Class Conditionals}

Finally, we perform an experiment similar to \cite{perezcruz} by studying the Chernoff information between the two's and three's, denoted $C(2,3)$, in the handwritten digit dataset MNIST \cite{mnist}. We apply a PCA with $d=32$ components to reduce the computational load and prevent overflow errors due to exponential scaling. We also plot a "null-consistency check"; by using samples only from the three class to report $C(3,3)$. Since they come from the same distribution, we expect $\widehat{C}(3,3) \approx 0$\footnote{Note here that the mixing parameter $s^*$ is not unique as the $C_s$ curve is flat and not strictly convex.}. \\

\begin{figure}
    \centering
    \includegraphics[width=\linewidth]{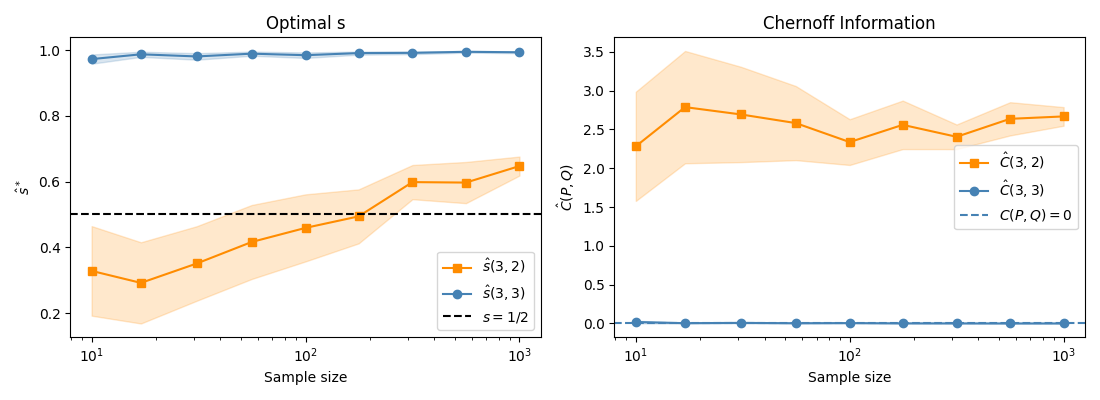}
    \caption{Chernoff information between MNIST digit classes as a function of per-class sample size $n=m$, estimated on $d=32$ PCA components. Orange curve is $\widehat{C}(2,3)$ between the two's and three's while blue is a null-consistency check, $\widehat{C}(3,3)$, computed from disjoint halves of the three samples. Bands show $95\%$ confidence intervals over $50$ repetitions. $\widehat{C}(3,3)$ stays near zero, while $\widehat{C}(2,3)$ is non-zero from as few as $10$ samples, so the estimator separates the two classes despite the small sample regime. Absolute $\widehat{C}(2,3)$ values are likely underestimates.}
\label{fig:mnist}
\end{figure}

Figure~\ref{fig:mnist} shows the $\widehat{C}(2,3)$ and $\widehat{C}(3,3)$ curves (as well as the mixing parameters $\hat{s}(2,3)$ and $\hat{s}(3,3)$) as a function of sample size. The reported $\widehat{C}(2,3)$ is likely far from the true value; the PCA projection can only reduce the divergence by the Data Processing Inequality \cite{polyanskiy}, while the finite-sample bias of the estimator adds an error of unknown sign. It is interesting to note that with as few as $10 - 50$ samples, the estimator yields a non-zero $\widehat{C}(2,3)$, while $\widehat{C}(3,3)$ stays around zero, suggesting that the estimator is able to distinguish between two sample sets easily. 

\section{Conclusion}

We introduced and analyzed a $k$-NN based estimator for Chernoff information, combining a derivative estimator with a bisection search over the mixing parameter $s \in (0,1)$. We then established the $L_1$- and $L_2$-consistency of the derivative estimator under mild regularity conditions on the underlying domain and the densities of the associated distributions, along with the convergence of the bisection to $s^*$ up to its precision parameter. \\

We further validated the estimator against analytically tractable ground truths using truncated exponential and truncated Gaussians on bounded domains. We further demonstrated its applicability to real high-dimensional data by distinguishing the class-conditional distributions of MNIST digits. \\

A notable limitation of the present analysis is that it is entirely asymptotic. No finite-sample guarantees or bias/variance trade-offs are provided, and the rate at which the estimator converges remains unknown. Moreover, the direction of the bias is unknown (i.e., whether it yields an under- or overestimate). Indeed, we argued that this prevents the use of a finite-difference estimator for the derivative and motivated the construction of $\widehat D_s$. In future work, we plan to derive the dependence of the bias and variance on $n$, $m$, $k$, and the smoothness of the underlying densities $p, q$ (similar to works such as \cite{zhao} and \cite{noshad} for other divergences).

\bibliographystyle{IEEEtran}
\bibliography{references}
\onecolumn
\appendices

\section{Proofs of Lemma \ref{lem:rmom}/\ref{lem:logmom}}\label{app:lemmaproofs}

Before we state the proofs, we define some basic notation and restate relevant results from \cite{PS}. Recall the $k$-NN volume ratio around a point $X_i$ is defined as
\begin{align}
  r_i \;:=\; \frac{(n-1)\,\rho_k(i)^d}{m\,\nu_k(i)^d},
\end{align}
with $\rho_k(i)$ the Euclidean distance from $X_i$ to its $k$-th nearest neighbour among $\{X_j\}_{j =1, j\ne i}^{n}$, and $\nu_k(i)$ the distance from $X_i$ to its $k$-th nearest neighbour among $\{Y_j\}_{j=1}^m$, for $n > k +1, m >k$. It is convenient to write
\begin{align}
  \zeta_n(x) := (n-1)\rho_k(1)^d, \qquad
  \theta_m(x) := m\nu_k(1)^d
\end{align}
so that, conditioning on $X_1 = x$,
\begin{align}\label{eq:ratio-decomp}
  r_1 =\ \frac{\zeta_n(x)}{\theta_m(x)}.
\end{align}
Write $\cbar$ for the volume of the unit ball in $\R^d$ and $\cB(x,R)$ for the closed ball of radius
$R$ about $x$. Further define the following quantities: 

\begin{align}
  \Pu := \int_{\cM\cap\cB(x,R_n(u))} p(t)\dd t,
  \qquad
  R_n(u) := \bigl(\tfrac{u}{n-1}\bigr)^{1/d}.
\end{align}

$\Pu$ corresponds to the amount of probability mass $p$ assigns on a region of volume $u$ around the point $x$. Note simply that $R_n(u)$ is the radius of the ball of volume $u$. Now we are ready to state the key lemma from \cite{PS} that determines the distribution of the scaled volume term $\zeta_n(x)$:

\begin{lemma}[Lemma 17 in \cite{PS}]\label{lem:cdf}
Conditional on $X_1=x$, the random variable $\zeta_n(x) = (n-1)\rho_k(1)^d$
has distribution function
\begin{align}\label{eq:F-formula}
  F_{n,k,x}(u) &= \PP\!\left(\zeta_n(x) < u \mid X_1 = x\right) \\
  &= 1 - \sum_{j=0}^{k-1}\binom{n-1}{j}\Pu^{\,j}\,
        \bigl(1-\Pu\bigr)^{n-1-j}
\end{align}

\end{lemma}

\begin{proofsketch}
    The event $\zeta_n(x) < u$ occurs exactly when \textbf{at least} $k$ of the points from $\{X_2, X_3, .., X_n\}$ fall inside the ball $\cB(x, R_n(u))$. Since each sample is taken independently, this corresponds to a Binomial experiment on $n-1$ points with success probability $\Pu$. The probability of at least $k$ successes is therefore equal to $F_{n,k,x}(u)$.
\end{proofsketch}

Next, we upper and lower bound $\Pu$. Two further quantities of interest are: 
\begin{align}
  \lambda_{\min} := c_p\,\rmin\,\cbar,
  \qquad
  \lambda_{\max} := C_p\,\cbar .
\end{align}

with $c_p, C_p, \rmin$ defined under Assumptions \ref{ass:dens} and \ref{ass:vol}.

\begin{lemma}[Boundedness of ball masses]\label{lem:ballmass} Under the standing assumptions \ref{ass:bdd} - \ref{ass:vol}, we have

\begin{enumerate}[label=\textup{(\roman*)}]
\item For all $u>0$, all $n$, and almost all $x\in\cM$,
\begin{align}\label{eq:Pupper}
  \Pu \;\le\; \lambda_{\max}\,\frac{u}{n-1}.
\end{align}
\item For all $u$ with $R_n(u)<1$ (equivalently, $u<n-1$), all $n$, and almost
all $x\in\cM$,
\begin{align}\label{eq:Plower}
  \Pu \;\ge\; \lambda_{\min}\,\frac{u}{n-1}.
\end{align}
\item For all $u$ with $R_n(u)\ge 1$ (equivalently, $u\ge n-1$), all $n$, and
almost all $x\in\cM$,
\begin{align}\label{eq:Pconst}
  \Pu \;\ge\; \lambda_{\min}.
\end{align}
\end{enumerate}
\end{lemma}

\begin{proof}
\emph{(i)} Since $p\le C_p$ and intersection only shrinks the ball,
\begin{align}
  \Pu \le C_p\,V(\cB(x,R_n(u))\cap\cM)
            &\le C_p\,V(\cB(x,R_n(u))) \\
            &= C_p\,\cbar\,R_n(u)^d \\
            &= \lambda_{\max}\,\frac{u}{n-1}.
\end{align}

\emph{(ii)} Since $p\ge c_p$ and $R_n(u)<1$, Assumption~\ref{ass:vol} applies
with $\delta=R_n(u)$, giving
$V(\cB(x,R_n(u))\cap\cM)\ge \rmin\cbar R_n(u)^d$, hence
$\Pu\ge c_p\rmin\cbar R_n(u)^d=\lambda_{\min}u/(n-1)$.\\

\emph{(iii)} For $R_n(u)\ge 1$ we have $\cB(x,1)\subseteq\cB(x,R_n(u))$, so
$V(\cB(x,R_n(u))\cap\cM)\ge V(\cB(x,1)\cap\cM)\ge\rmin\cbar$ by
Assumption~\ref{ass:vol}, and the bound follows on multiplying by $c_p$.
\end{proof}

Finally, we state two elementary facts without proof, to be used throughout.
\begin{lemma}\label{lem:elem}
  The following hold
  \begin{enumerate}[label=(\alph*)]
    \item For $p\in[0,1]$, $1-p\le e^{-p}$.
    \item For $0\le j\le n-1$ and $n > 1, n \in \mathbb{N}$, $\dbinom{n-1}{j}\le \dfrac{(n-1)^j}{j!}$
  \end{enumerate}
\end{lemma}

\subsection{Positive Moments of $\zeta_n(x)$}

\begin{lemma}\label{lem:posmom}
For any fixed $t>0$,
\begin{align}
\esssup_{x\in\cM}\,\sup_{n>k}\,\E\left[\zeta_n(x)^{t}\mid X_1=x\right]
  < \infty.
\end{align}
\end{lemma}

\begin{proof}
    By  the Layer-Cake Identity \cite{lieb} (see also Lemma~31 of \cite{PS}) we have,
\begin{align}\label{eq:pos-layer}
  \E\left[\zeta_n(x)^t\right]
  = t\int_0^\infty u^{t-1}\bigl(1-F_{n,k,x}(u)\bigr)\dd u .
\end{align}
First, we treat the small $n$ regime where $k < n < 2k$:\\

For each such $n$, Assumption~\ref{ass:bdd} gives $\rho_k(1)\le D$ almost surely, so
$\zeta_n(x)\le (n-1)D^d \le (2k-1)D^d$ almost surely. Hence $\E[\zeta_n(x)^t]\le((2k-1)D^d)^t$, independent of $x$. Since there are only finitely many such $n$, this contributes a finite supremum to the overall bound. Therefore, for the remainder of the proof, assume $n\ge 2k$, so that $(n-1-j)/(n-1)\ge 1/2$ for every $j\le k-1$.\\

We now separate the expectation into three regions on $u$:\\

\textbf{Case $\boldsymbol{0<u<n-1}$:}
Both bounds of Lemma~\ref{lem:ballmass} apply. For each $j\le k-1$,
\begin{align}
  \binom{n-1}{j}\Pu^{j}
  &\le \frac{(n-1)^j}{j!}\cdot \Pu^j
  \tag{Lem.~\ref{lem:elem}(b)}\\
  &\le \frac{(n-1)^j}{j!}\cdot\left(\frac{\lambda_{\max}u}{n-1}\right)^j
  \tag{\eqref{eq:Pupper}}\\
  &= \frac{(\lambda_{\max}u)^{j}}{j!},
\end{align}
and
\begin{align}
  (1-\Pu)^{n-1-j}
  &\le \exp\bigl(-(n-1-j)\,\Pu\bigr)
  \tag{Lem.~\ref{lem:elem}(a)}\\
  &\le \exp\left(-(n-1-j)\,\frac{\lambda_{\min}u}{n-1}\right)
  \tag{\eqref{eq:Plower}}\\
  &= \exp\left(-\frac{n-1-j}{n-1}\,\lambda_{\min}u\right)\\
  &\le e^{-\lambda_{\min}u/2},
  \tag{$n\ge 2k$}
\end{align}
Summing over $j$ and applying~\eqref{eq:F-formula},
\begin{align}\label{eq:Range1bound}
  1-F_{n,k,x}(u) \le
  \Biggl(\sum_{j=0}^{k-1}\frac{(\lambda_{\max}u)^{j}}{j!}\Biggr)
  e^{-\lambda_{\min}u/2},
  \ 0<u<n-1.
\end{align}

Plugging back to the layer-cake identity, we get:
\begin{align}
 & t\int_0^{n-1} u^{t-1}\bigl(1-F_{n,k,x}(u)\bigr)\,\dd u \\
  &\le t\int_0^{\infty} u^{t-1}
  \Biggl(\sum_{j=0}^{k-1}\frac{(\lambda_{\max}u)^{j}}{j!}\Biggr)
  e^{-\lambda_{\min}u/2}\,\dd u\\
  &= t\sum_{j=0}^{k-1}\frac{\lambda_{\max}^j}{j!}
  \int_0^{\infty} u^{t+j-1}\,e^{-\lambda_{\min}u/2}\,\dd u,
\end{align}
Each term in the sum is a polynomial-times-exponential integrand $u^{t+j-1}e^{-\lambda_{\min}u/2}$; since exponential decay dominates any polynomial growth, every such integral is finite and their finite sum is finite. \\

\textbf{Case $\boldsymbol{n-1\le u\le (n-1)D^d}$:}
For $u\ge n-1$, Lemma~\ref{lem:ballmass}(iii) gives
$\Pu\ge\lambda_{\min}$. For each $j\le k-1$,
\begin{align}
  \binom{n-1}{j}\Pu^j
  &\le \frac{(n-1)^j}{j!}\cdot 1
  \tag{Lem.~\ref{lem:elem}(b), $P\le 1$}\\
  &\le \frac{(n-1)^{k-1}}{(k-1)!},
  \tag{$j\le k-1 < n -1 $}
\end{align}
and
\begin{align}
  (1-\Pu)^{n-1-j}
  &\le \exp \bigl(-(n-1-j)\,\Pu\bigr)
  \tag{Lem.~\ref{lem:elem}(a)}\\
  &\le e^{-(n-1-j)\lambda_{\min}}
  \tag{\eqref{eq:Pconst}}\\
  &\le e^{-(n-k)\lambda_{\min}}
  \tag{$j\le k-1$}\\
  &\le e^{-\lambda_{\min}n/2}.
  \tag{$n\ge 2k$}
\end{align}
Summing over $j\le k-1$ and applying~\eqref{eq:F-formula},
\begin{align}
  1-F_{n,k,x}(u)
  \le k\cdot\frac{(n-1)^{k-1}}{(k-1)!}\,e^{-\lambda_{\min}n/2}.
\end{align}
The right-hand side is independent of $u$, so the contribution of this range to~\eqref{eq:pos-layer} is bounded by
\begin{align}
  &t\int_{n-1}^{(n-1)D^d} u^{t-1}\cdot
  k\cdot\frac{(n-1)^{k-1}}{(k-1)!}\,e^{-\lambda_{\min}n/2}\,\dd u \\
  &= \frac{k(n-1)^{k-1}}{(k-1)!}\,e^{-\lambda_{\min}n/2}
  \cdot\Bigl[(n-1)^t D^{dt} - (n-1)^t\Bigr]\\
  &\le \frac{k\,D^{dt}}{(k-1)!}\,(n-1)^{t+k-1}\,e^{-\lambda_{\min}n/2}.
\end{align}
Similar to the previous case, this is a polynomial in $n$ multiplied by a decaying exponential in $n$;
exponential decay dominates polynomial growth, so the supremum is finite.\\

\textbf{Case $\boldsymbol{u>(n-1)D^d}$:}
Here $R_n(u)>D$, so the ball is larger than the domain; $\mathcal{B}(x,R_n(u))\supseteq\mathcal{M}$
and therefore $\Pu=\int_{\mathcal{M}}p(t)\,\dd t=1$, hence $1-F_{n,k,x}(u)=0$ and this range contributes
nothing to~\eqref{eq:pos-layer}.\\

Combining all cases plus the small $n$ range yields a finite supremum over $n,x$.
\end{proof}

\subsection{Negative Moments of $\zeta_n(x)$}
\begin{lemma}\label{lem:negmom}
For any fixed $t$ with $0<t<k$,
\begin{align}
  \esssup_{x\in\cM}\,\sup_{n>k}\,\E\left[\zeta_n(x)^{-t}\mid X_1=x\right]
  < \infty.
\end{align}
\end{lemma}

\begin{proof}
By the Layer-Cake Identity \cite{lieb} (see also Lemma~32 in \cite{PS}), for $t>0$,
\begin{align}
  \E\left[\zeta_n(x)^{-t}\right]
  = t\int_0^\infty u^{-t-1}\,F_{n,k,x}(u)\dd u .
\end{align}
We can rewrite \eqref{eq:F-formula} as follows (using the "flipped" counting argument)
\begin{align}
  F_{n,k,x}(u)
  = \sum_{j=k}^{n-1}\binom{n-1}{j}\Pu^{j}(1-\Pu)^{n-1-j}
\end{align}

Here we need to only distinguish two cases for $u$ \\

\textbf{Case $\boldsymbol{u>1}$:}
Using only $F_{n,k,x}(u)\le 1$,
\begin{align}
  t\int_{1}^{\infty}u^{-t-1}F_{n,k,x}(u)\,\dd u
  &\le t\int_{1}^{\infty}u^{-t-1}\dd u \\
  &= t\left[\frac{u^{-t}}{-t}\right]_{1}^{\infty}
  = t\cdot\frac{1}{t}
  = 1,
\end{align} for $t \in (0, k)$. Intuitively, this corresponds to the "unproblematic" region where the negative moment can not get arbitrarily large since the ball volume is already large enough (whose inverse powers will be small). \\

\textbf{Case $\boldsymbol{0<u\le 1}$:}
Dropping the factors $(1-\Pu)^{n-1-j}\le 1$ in~\eqref{eq:F-formula},
\begin{align}
  F_{n,k,x}(u)
  &= \sum_{j=k}^{n-1}\binom{n-1}{j}\Pu^j(1-\Pu)^{n-1-j} \\
  &\le \sum_{j=k}^{n-1}\binom{n-1}{j}\Pu^j
  \le \sum_{j=k}^{n-1}\frac{(\lambda_{\max}u)^j}{j!},
\end{align}
where the last step follows from Lemma~\ref{lem:elem}(b) and~\eqref{eq:Pupper}. Since each term is positive, we extend the sum to $n\to \infty$ and pull the $u^k$ term out
\begin{align}
  F_{n,k,x}(u) 
  \le \sum_{j=k}^{\infty}\frac{(\lambda_{\max}u)^j}{j!} = u^k \sum_{j=k}^{\infty}\frac{\lambda_{\max}^j u^{j-k}}{j!} 
\end{align}
Noting that $u \leq 1 \implies u^{j-k} \leq 1$ we get
\begin{align}
  F_{n,k,x}(u) 
  \le u^k \sum_{j=k}^{\infty}\frac{\lambda_{\max}^j}{j!} \le u^k e^{\lambda_{\max}}, 
\end{align} by the power series expansion of $e^x$. Hence
\begin{align}
  t\int_0^1 u^{-t-1}F_{n,k,x}(u)\,\dd u
  &\le e^{\lambda_{\max}} t \int_0^1 u^{k-t-1}\,\dd u \\
  &= \frac{e^{\lambda_{\max}}t}{k-t}, \ \text{for $t \in (0,k).$}
\end{align} 
Adding the two parts yields the claim. Note that the bound depends only on $t,k$ and $\lambda_{\max}$.
\end{proof}

\begin{remark}
The integrability of the small $u$ region introduces the constraint on $k$, since for $k < t$, the integrand gets arbitrarily large as $u \to 0$.
\end{remark}

\begin{remark}
The negative-moment proof uses only the upper bound on $\Pu$ (which is free from $V(\cB\cap\cM)\le V(\cB)$). Neither Assumption~\ref{ass:vol} nor Assumption~\ref{ass:bdd} is invoked. Both are used only in the positive-moment proof but we state the lemma with all for symmetry and readability.
\end{remark}

Finally, note that by symmetry, propositions ~\ref{lem:posmom} and~\ref{lem:negmom} apply equally
to $\theta_m(x)$ (with $p$ replaced by $q$, the constants $(c_p,C_p)\mapsto(c_q,C_q)$ and the binomial distributions Bin$(n-1, \Pu) \mapsto $ Bin$(m, Q_{m,u,x})$ for $m \geq k$). We now have all the needed intermediate results to state the  proofs.

\subsection{Proof of Lemma~\ref{lem:rmom}}
Recall the claim, $\forall |\beta|<k$, we have
\begin{align}
  \sup_{n>k,m\geq k,i}\,\E\left[\,r_i^{\,\beta}\,\right] < \infty .
\end{align} 

Since all samples are i.i,d, we take $i=1$. Conditioning on $X_1=x$ and using \eqref{eq:ratio-decomp} together with the conditional independence of
$\zeta_n(x)$ (a function of the $X$-sample) and $\theta_m(x)$ (a function of
the $Y$-sample),
\begin{align}
  &\E\left[r_1^{\,\beta}\mid X_1=x\right] \\
  =& \E\left[\zeta_n(x)^{\beta}\mid X_1=x\right]\cdot
    \E\left[\theta_m(x)^{-\beta}\mid X_1=x\right]
\end{align}
If $\beta>0$, both factors are finite uniformly: the first by Lemma ~\ref{lem:posmom}, the second by Lemma~\ref{lem:negmom} (applied to $\theta_m$ at exponent $\beta<k$). If $\beta<0$, swap the roles; the first factor is a negative moment of $\zeta_n$ of order $|\beta|<k$, finite by Lemma~\ref{lem:negmom}, and the second is a positive moment of $\theta_m$, finite by Lemma~\ref{lem:posmom}.

\subsection{Proof of Lemma~\ref{lem:logmom}}
Recall the claim,  $\forall |\beta|<k, \theta\ge 0$, we have

\begin{align}
  \sup_{n,m,i}\,\E\left[\,r_i^{\,\beta}\,\bigl|\log r_i\bigr|^{\theta}\,\right]
  < \infty .
\end{align}

For every $\theta\ge 0$ and every $\eta>0$ there is a constant $C_{\theta,\eta}<\infty$
with
\begin{align}\label{eq:logpoly}
  |\log v|^{\theta} \le C_{\theta,\eta}\bigl(v^{\eta}+v^{-\eta}\bigr),
  \qquad v>0.
\end{align}
\textbf{For $v\ge 1$:} $(\log v)^\theta/v^\eta\to 0$ as $v\to\infty$, so the ratio is
bounded on $[1,\infty)$. \\

\textbf{For $0<v\le 1$:} substitute $w=1/v$ to reduce to the previous case. \\

In both cases the bound is uniform in $v$, with constant depending only on $\theta$ and $\eta$, so set the maximum of the two to be $C_{\theta,\eta}$.\\

Apply \eqref{eq:logpoly} with $v=r_i$ and an $\eta>0$ chosen small enough that $|\beta|+\eta<k$, which is possible since $|\beta|<k$ strictly. Then
\begin{align}
  r_i^{\,\beta}\bigl|\log r_i\bigr|^{\theta}
  \le C_{\theta,\eta}\bigl(r_i^{\,\beta+\eta}+r_i^{\,\beta-\eta}\bigr),
\end{align}
and $|\beta\pm\eta|\le|\beta|+\eta<k$. Taking expectations and applying Lemma~\ref{lem:rmom} to each term concludes the proof.

\section{Proofs of Theorem \ref{thm:L1}/\ref{thm:L2}}\label{app:thmproofs}

\subsection{Limits of $\E[A_n] \ \text{and} \ \E[B_n]$}
We begin by defining the relevant quantities again:
\begin{align}
    A_n &= \frac{1}{n} \sum_{i=1}^n r_i^{\,\gamma} \quad B_n = \frac{1}{n} \sum_{i=1}^n r_i^{\,\gamma} \log r_i.\\
    a_s &= \Binv C_s(P,Q), \quad b_s = \Binv C_s(P,Q)\big(G_{k,s} - D_s(P,Q) \big) 
\end{align}
with $\widehat D_s = G_{k,s} - B_n/A_n$. As mentioned before, the consistency of $A_n$ is proven in \cite{PS}, which we state now for completeness.

\begin{lemma}[Theorem 6 in \cite{PS}]\label{lem:EAn} For the estimator $A_n$ and $k > 1$ we have
\begin{align}
\lim_{n,m\to \infty} \E [A_n(X^n, Y^m)] = \Binv C_s(P,Q) = a_s
\end{align}
\end{lemma}

\begin{remark}
    One needs to check that our assumptions are as strong as that of the assumptions in \cite{PS}. Indeed we have the following: 
    
    \begin{itemize}
        \item[(a)] Satisfied by choice of $\gamma = 1-s \in (0,k)$.
        \item[(b)] Satisfied by Assumption~\ref{ass:dens}: $p \geq c_p > 0$.
        \item[(c)] Uniform Lebesgue approximability is used to bound the kNN ball volumes; an equivalent formulation is proved in Lemma~\ref{lem:ballmass} following again from Assumption~\ref{ass:dens}.
        \item[(d)] Follows from Assumptions~\ref{ass:bdd} and~\ref{ass:dens}: boundedness of $\mathcal{M}$ and $p \leq C_p$ give uniform control over the integral of $H(x,p,\delta,1)$.
        \item[(e)] Follows from Assumption~\ref{ass:bdd}: $\|x-y\|^\gamma \leq D^\gamma < \infty$ uniformly, so $\int_\mathcal{M} \|x-y\|^\gamma p(y)\,dy \leq D^\gamma < \infty$ for all $x$.
        \item[(f)] Same argument as (e) by Fubini's Theorem \cite{lieb}.
        \item[(g)] Satisfied by Assumption~\ref{ass:dens}: $q \leq C_q < \infty$.
    \end{itemize}
\end{remark}

The following lemma states the consistency of $B_n$:
\begin{lemma}\label{lem:EBn}
    For the estimator $B_n$ and $k \ge 5$ we have
    \begin{align}
        \lim_{n,m \to \infty} \E[B_n(X^n, Y^m)] =
        \Binv C_s(P,Q)\big(G_{k,s} - D_s(P,Q) \big)  = b_s
    \end{align}
\end{lemma}
\begin{proof}
    Writing the expectation explicitly and using that $X_i$ are i.i.d, we condition on $X_1 = x$ and define the following four functions,
        \begin{align}
        f_n(x) &:= \E[\zeta_n^\gamma \mid X_1 = x],  \quad
        \widetilde f_n(x) := \E[\zeta_n^\gamma \log \zeta_n \mid X_1 = x], \\
        g_m(x) &:= \E[\theta_m^{-\gamma} \mid X_1 = x],  \
        \widetilde g_m(x) := \E[\theta_m^{-\gamma} \log \theta_m | X_1 = x].
    \end{align}
Here $f_n, \widetilde f_n$ are the $\gamma$-moment and log-weighted $\gamma$-moment of the $p$-sample $k$-NN volumes and similarly $g_m, \widetilde g_m$ the $(-\gamma)$-moment and log-weighted $(-\gamma)$-moment of the $q$-sample. The conditional expectation is
    \begin{align}
        \E[r_1^\gamma \log r_1 \mid X_1 = x] = \widetilde f_n(x)\, g_m(x) - \widetilde g_m(x)\, f_n(x),
    \end{align}
    and integrating against the law of $X_1$, for any $n > k$,
    \begin{align} \label{eqn:EBn}
        \E[B_n] = \int_{\mathcal{M}} \left( \widetilde f_n(x)\, g_m(x) - \widetilde g_m(x)\, f_n(x) \right) p(x)\, dx.
    \end{align}

The finiteness of each term is given by Lemmas 1/2 but we need to identify the exact values now. For fixed $n,m$, Lemma~\ref{lem:cdf} gives the CDF of each volume term. This CDF is known to converge in law to Erlang variables, and the moments of Erlang variables are known in closed form, so the following lemma from \cite{PS}, stated without proof, combines these results.

\begin{lemma}\label{lem:erlang}[Lemma 16/18 in \cite{PS}, also see \cite{leonenko}]
    For almost all $x \in \cM, F_{n,k,x} \to^n_{weakly} F_{k,x}$ where $F_{k,x}$ is the Erlang variable with parameter $\lambda(x) = \bar{c}p(x)$ and cumulative distribution function:
    \begin{align}
        F_{k,x}(u) = 1 - \exp(-\lambda u)\sum_{j=0}^{k-1} \frac{(\lambda u)^j}{j!}
    \end{align}
    and density 
    \begin{align}
        f_{x,k}(u) = \frac{1}{\Gamma(k)} \lambda^k(x) u^{k-1}\exp(-\lambda(x)u)
    \end{align}

    Moreover for any $\beta \in \R$ (with $\beta + k >0$) it has $\beta$'th moment:
    \begin{align}
        M_x(\beta) = \int_0^\infty u^\beta f_{x,k}(u) du = \lambda(x)^{-\beta}\frac{\Gamma(k + \beta)}{\Gamma(k)}
    \end{align}
\end{lemma}

Weak convergence from Lemma~\ref{lem:erlang} does not by itself give convergence in moments; $\E[\zeta_n(x)^\gamma \mid X_1 = x] \to \E[\zeta(x)^\gamma]$. Lemma~\ref{lem:momconv} establishes this via uniform integrability.

\begin{lemma}[Theorem~20/21 in \cite{PS}]\label{lem:momconv}
Let $k + \gamma > 0$. Then for almost all $x \in \cM$,
\begin{align}
    \lim_{n \to \infty} f_n(x) = \E[\zeta(x)^\gamma] = M_x^p(\gamma) = (\bar c\, p(x))^{-\gamma}\,\frac{\Gamma(k+\gamma)}{\Gamma(k)},
\end{align}
where $\zeta(x) \sim \mathrm{Erlang}(k, \bar c\, p(x))$ is the weak limit of $\zeta_n(x)$. Likewise, if $k - \gamma > 0$, then for almost all $x \in \cM$,
\begin{align}
    \lim_{m \to \infty} g_m(x) = \E[\theta(x)^{-\gamma}] = M_x^q(-\gamma) = (\bar c\, q(x))^{\gamma}\,\frac{\Gamma(k-\gamma)}{\Gamma(k)},
\end{align}
where $\theta(x) \sim \mathrm{Erlang}(k, \bar c\, q(x))$ is the weak limit of $\theta_m(x)$.
\end{lemma}

\begin{proof}
By Lemma~\ref{lem:erlang}, $\zeta_n(x) \to_d \zeta(x)$ with $\zeta(x) \sim \mathrm{Erlang}(k, \bar c\, p(x))$. Since $u \mapsto u^\gamma$ is continuous on $(0,\infty)$ and $\zeta_n(x) \in (0,\infty)$ almost surely, the Continuous Mapping theorem \cite{dervaart} gives $\zeta_n(x)^\gamma \to_d \zeta(x)^\gamma$. Convergence in distribution does not transfer to expectations in general, so we invoke the limit-of-moments criterion (Lemma~12 in \cite{PS}, after van der Vaart \cite{dervaart}): if $Z_n \to_d Z$ with $Z_n, Z \ge 0$ and there exists $\epsilon > 0$ with $\limsup_n \E[Z_n^{1+\epsilon}] < \infty$, then $\E[Z_n] \to \E[Z]$. Applying this to $Z_n = \zeta_n(x)^\gamma$, it suffices to exhibit an $\epsilon > 0$ with $k + \gamma(1+\epsilon) > 0$ and
\begin{align}
    \limsup_{n \to \infty} \E[\zeta_n(x)^{\gamma(1+\epsilon)} \mid X_1 = x] < \infty \quad \text{for a.a.\ } x \in \cM,
\end{align}
which directly follows from Lemma~\ref{lem:posmom}. Hence $f_n(x) = \E[\zeta_n(x)^\gamma \mid X_1 = x] \to \E[\zeta(x)^\gamma]$, and Lemma~\ref{lem:erlang} evaluates the limit as $M_x^p(\gamma)$. The statement for $g_m$ is identical: $\theta_m(x) \to_d \theta(x)$, apply the criterion with exponent $-\gamma$ (requiring $k - \gamma > 0$ and $\epsilon$ small enough that $k - \gamma(1+\epsilon) > 0$, then Lemma~\ref{lem:negmom} suffices), and evaluate the $(-\gamma)$-moment via Lemma~\ref{lem:erlang}.
\end{proof}

For the log terms $\widetilde{f_n}$ and $\widetilde{g_m}$, start by simply noting
\begin{align}
    \frac{\partial}{\partial \beta} u^\beta f_{x,k}(u)=  u^\beta \log u \, f_{x,k}(u)
\end{align}

Thus, if we could justify moving the partial derivative inside the expectation, we can compute the values of $\widetilde{f_n}, \widetilde{g_m}$ by an argument identical to above. The following lemma establishes the justification via dominated convergence \cite{lieb}. The proof follows a similar structure to Lemma~\ref{lem:logmom}.

\begin{lemma}\label{lem:erlanglog}
For almost all $x \in \cM$ and $\beta \in \R,$ s.t $ \beta + k > 0$, we have
\begin{align}
    \frac{\partial}{\partial \beta} M_x(\beta) 
    &= \lambda(x)^{-\beta}\frac{\Gamma(k+\beta)}{\Gamma(k)}\big(\psi(k+\beta)-\log \lambda(x)\big). \\
    &=\E[u^\beta \log u \mid X_1 = x] 
\end{align}    
\end{lemma}
\begin{proof}
Fix $\beta$ in a neighborhood of $\gamma$, e.g. $\beta\in[\gamma/2,3\gamma/2]$. We construct an integrable envelope in this interval by considering the cases when $\log u > 0$ and $\log u < 0$ separately. \\

\textbf{Case \boldmath$0<u<1$:} Since $\beta\ge \gamma/2$, we have $u^\beta \le u^{\gamma/2}$ on $(0,1)$, hence
\begin{align}
u^\beta |\log u| \le u^{\gamma/2}|\log u| = u^{\gamma/4}\bigl(u^{\gamma/4}|\log u|\bigr) \le C_1\, u^{\gamma/4},
\end{align}
where $C_1 = \sup_{u\in(0,1)} u^{\gamma/4}|\log u| = \frac{4}{e\gamma}$.\\

\textbf{Case \boldmath$u>1$:} Since $\beta\le 3\gamma/2$,
\begin{align}
u^\beta |\log u| \le u^{3\gamma/2}\log u \le C_2 u^{3\gamma/2+1},
\end{align}
where $C_2 = \sup_{u>1} u^{-1}\log u = \frac{1}{e}$.
Hence, for $C:=\max\{C_1,C_2\}$,
\begin{align}
|u^\beta \log u| \le
C\Big(u^{\gamma/4}\mathbf 1_{(0,1)}(u)+u^{3\gamma/2+1}\mathbf 1_{(1,\infty)}(u)\Big).
\end{align}

For completeness, at $u=1$ note simply that $u|\log u| = 0$. Hence, we can define the envelope, with $f_{x,k}(u)$ the Erlang density from Lemma~\ref{lem:erlang}
\begin{align}
g_x(u) := C\Big(u^{\gamma/4}\mathbf 1_{(0,1)}(u)+u^{3\gamma/2+1}\mathbf 1_{(1,\infty)}(u)\Big)f_{x,k}(u).
\end{align}
For $0 < u < 1$, $g_x(u)$ behaves like $u^{k-1+\gamma/4}$ (which is integrable for $k>1$), and for $u>1$ the exponential decay of $f_{x,k}(u)$ ensures integrability. Hence we have $g_x\in L^1(0,\infty)$ and by Dominated Convergence \cite{lieb},
\begin{align}
\frac{d}{d\beta}M_x(\beta) =
\int_0^\infty u^\beta \log u\, f_{x,k}(u)\,du = \mathbb E[\zeta(x)^\beta\log \zeta(x)].
\end{align}
Differentiating $M_x(\beta)$ and evaluating at $\beta=\gamma$ yields
\begin{align}
\mathbb E[\zeta(x)^\gamma\log \zeta(x)] = \lambda(x)^{-\gamma}\frac{\Gamma(k+\gamma)}{\Gamma(k)}\big(\psi(k+\gamma)-\log \lambda(x)\big).
\end{align}
\end{proof}
As an immediate corollary we get
\begin{corollary}\label{cor:logmom} For almost all $x \in \cM$ and $\gamma$ s.t. $k + \gamma > 0$ and $k - \gamma > 0$
\begin{align}
\lim_{n \to \infty} \widetilde f_n(x) &= 
(\bar{c} p(x))^{-\gamma}
\frac{\Gamma(k+\gamma)}{\Gamma(k)}
\big(\psi(k+\gamma)-\log(\bar{c} p(x))\big), \\
\lim_{n \to \infty} \widetilde g_m(x) &=
(\bar{c} q(x))^{\gamma}
\frac{\Gamma(k-\gamma)}{\Gamma(k)}
\big(\psi(k-\gamma)-\log(\bar{c} q(x))\big).
\end{align}
\end{corollary}

Thus the limit of each conditional expectation is identified. The final step is to justify switching the limit and the expectation. However, since the conditional moments are already established to be pointwise bounded by Lemmas~\ref{lem:posmom}-~\ref{lem:negmom}, Dominated Convergence of $\E[B_n]$ follows from a constant envelope. The next lemma makes this argument precise. 

\begin{lemma}\label{lem:EBnlimit}
Under the standing assumptions \ref{ass:bdd} - \ref{ass:vol}, for $k \ge 5$, we have
\begin{align}
\lim_{n,m\to\infty}\E[B_n]
= \int_{\cM}\Bigl(&\lim_{n\to\infty}\widetilde{f}_n(x)\,\lim_{m\to\infty}g_m(x)\\
- &\lim_{m\to\infty}\widetilde{g}_m(x)\,\lim_{n\to\infty}f_n(x)\Bigr)\,p(x)\,dx,
\end{align}
the four inner limits being those of Lemma~\ref{lem:momconv} and Corollary~\ref{cor:logmom}.
\end{lemma}

\begin{proof}
By Lemma~\ref{lem:momconv} and Corollary~\ref{cor:logmom} the integrand $\widetilde{f}_n g_m - \widetilde{g}_m f_n$ converges pointwise for almost all $x\in\cM$, so it suffices to dominate it by a constant. Fix $\eta\in(0,k-\gamma)$, (the interval is non-empty since $k \ge 5$ and $\gamma \in (0,1)$)) and apply the bound $|\log u|\le C_\eta(u^{\eta}+u^{-\eta})$ from Lemma~\ref{lem:logmom}:
\begin{align}
|\widetilde{f}_n(x)|
&\le \E\!\left[\zeta_n(x)^{\gamma}|\log\zeta_n(x)| \mid X_1=x\right] \\
&\le C_\eta\Bigl(\E\!\left[\zeta_n(x)^{\gamma+\eta}\mid X_1=x\right] +\E\!\left[\zeta_n(x)^{\gamma-\eta}\mid X_1=x\right]\Bigr).
\end{align}
Both exponents satisfy $|\gamma\pm\eta|<k$, so Lemmas~\ref{lem:posmom} and~\ref{lem:negmom} bound each term uniformly in $(n,x)$; the same estimate with $q$ in place of $p$ bounds $\widetilde{g}_m$, and $f_n,g_m$ are bounded directly by Lemmas~\ref{lem:posmom}--\ref{lem:negmom}. Hence
\begin{align}
\sup_{n,m>k}\ \sup_{x\in\cM}\ \bigl|\widetilde{f}_n(x)\,g_m(x)-\widetilde{g}_m(x)\,f_n(x)\bigr| =: K < \infty.
\end{align}
Since $p(x)\,dx$ is a probability measure, we have $\int_{\cM} Kp(x)dx < \infty $ and Dominated Convergence gives the claim.
\end{proof}

Simply plugging Corollary~\ref{cor:logmom} and Lemma~\ref{lem:momconv}, we thus conclude:

\begin{align}
\widetilde f_n g_m-\widetilde g_m f_n
\longrightarrow
B_{k,s}^{-1}\Bigl(\tfrac{q(x)}{p(x)}\Bigr)^{\!\gamma}
   \Bigl(G_{k,s}+\log\tfrac{q(x)}{p(x)}\Bigr),
\end{align}
where
\begin{align}
B_{k,s}^{-1}=\frac{\Gamma(k+\gamma)\Gamma(k-\gamma)}{\Gamma(k)^2},
\quad
G_{k,s}=\psi(k+\gamma)-\psi(k-\gamma),
\end{align}
with the factor $\cbar$ cancelling. Lemma~\ref{lem:EBnlimit}
justifies passing the limit through the integral, so
\begin{align}
&\lim_{n,m\to\infty}\E[B_n] = B_{k,s}^{-1}\int_{\cM}\left(\frac{q(x)}{p(x)}\right)^{\!\gamma}
   \left(G_{k,s}+\log\frac{q(x)}{p(x)}\right) p(x)\,d\mu(x).
\end{align}
The two integrals evaluate through
\begin{align}
&G_{k,s}\int_{\cM}\left(\frac{q(x)}{p(x)}\right)^{\!\gamma} p(x)\,d\mu(x) =G_{k,s}\,C_s(P,Q),\\
&\int_{\cM}\left(\frac{q(x)}{p(x)}\right)^{\!\gamma}\log\frac{q(x)}{p(x)} p(x)\,d\mu(x)
= -\partial_s C_s(P,Q) = -\,C_s(P,Q)\,D_s(P,Q),
\end{align}
the second using $\partial_s C_s(P,Q)=\int_{\cM}p^{s}q^{1-s}\log(p/q)\,d\mu$. Hence
\begin{align}
\lim_{n,m\to\infty}\E[B_n]
= B_{k,s}^{-1}\,C_s(P,Q)\bigl(G_{k,s}-D_s(P,Q)\bigr) = b_s.
\end{align}
\end{proof}


\subsection{Convergence of the Ratio}
As stated previously, Lemmas~\ref{lem:EAn} and \ref{lem:EBn} show only the convergence of the individual terms. We now prove that the ratio of the two also converges. A sketch of these steps is given in Section~\ref{sec:L1}. The first lemma taken from \cite{PS} establishes the variance decay of $A_n$.
\begin{lemma}[Theorem 8 in \cite{PS}]\label{lem:varAn}
\begin{align}
    \lim_{n,m \to \infty} \Var(A_n) = 0
\end{align}

\end{lemma}
\begin{remark}
Again, we need to check that the assumptions from \cite{PS} are implied by our standing assumptions. 
\begin{enumerate}[label=(\alph*)]
    \item Satisfied by choice of $\gamma = 1-s \in (0, (k-1)/2)$,
    the sharper range now binding through the covariance terms; met for
    $\gamma \in (0,1)$ by $k \ge 5$.
    \item Satisfied by Assumption~\ref{ass:dens}: $p \geq c_p > 0$.
    \item Uniform Lebesgue approximability, following from Assumption~\ref{ass:dens}.
    \item Follows from Assumptions~\ref{ass:bdd} and~\ref{ass:dens}: boundedness of $\mathcal{M}$
    and $p \leq C_p$ give uniform control over the integral of
    $H(x, p, \delta, 1/2)$, now at $\omega = 1/2$.
    \item Follows from Assumption 1: $\|x-y\|^\gamma \leq D^\gamma < \infty$
    uniformly, so $\int_{\cM} \|x-y\|^\gamma p(y)\, dy \leq D^\gamma < \infty$
    for all $x$.
    \item Same argument as (e) by Fubini's Theorem \cite{lieb}.
    \item Satisfied by Assumption~\ref{ass:dens}: $q \leq C_q < \infty$.
\end{enumerate}
\end{remark}

Following similar steps, we prove that $\Var (B_n) \to 0$.

\begin{lemma}\label{lem:varBn}
For $k \ge 5$, we have
\begin{align} 
    \lim_{n,m \to \infty} \Var(B_n) = \lim_{n,m \to \infty} \E[(B_n - b_s)^2] = 0
\end{align}

\end{lemma}
\begin{proof}
The first observation is that one can treat the diagonal and off-diagonal expectations separately, write
\begin{align}
B_n = \frac{1}{n}\sum_{i=1}^n W_i,
\qquad
W_i := r_i^\gamma \log r_i.
\end{align}
Using that $W_i$ are identically distributed, for the quadratic term we have:
\begin{align}
\mathbb E[B_n^2]
&=
\mathbb E\!\left[\left(\frac{1}{n}\sum_{i=1}^n W_i\right)^2\right]
=
\frac{1}{n^2}\mathbb E\!\left[\sum_{i=1}^n W_i^2 + \sum_{i\neq j} W_i W_j \right] \\
&= \frac{1}{n}\mathbb E[W_1^2] +
\frac{n-1}{n}\mathbb E[W_1 W_2].
\end{align}
By Lemma~\ref{lem:logmom}, $\E[W_1^2]$ is finite and therefore $\lim_{n\to\infty} 1/n \,\E[W_1^2]$ vanishes. For the cross term $\E[W_1W_2]$, conditioning on $(X_1, X_2) = (x_1, x_2)$ we distinguish two cases: \\

\textbf{(a) Overlapping Balls:} Let $R_n(u) = (\frac{u}{n-1})^{1/d}$ and $R_n(v) = (\frac{v}{n-1})^{1/d}$ be the radii of the two balls. Then for fixed $u,v$ and $x_1\neq x_2$:
\begin{align}
&\mathbb P\big(\cB(x_1,R_n(u)) \cap \cB(x_2,R_n(v)) \neq \varnothing\big)\\ 
=& 
\mathbb P\big(R_n(u)+R_n(v) \ge \|x_1-x_2\|\big)
\end{align}
since $(R_n(u)+R_n(v)) \to 0$, the contribution from overlapping configurations also vanish.\\

\textbf{(b) Disjoint Balls:} In this case, each sample point $X_i$ is either inside $\cB(x_1,R_n(u)), \cB(x_2,R_n(v))$ or outside of both balls. Define the joint distribution function, analogously to Lemma~\ref{lem:cdf}, as
\begin{align}
&\mathbb P\!\left(
\zeta_{n,k}(x_1)<u \land \zeta_{n,k}(x_2)<v
\,\middle|\,
X_1=x_1,X_2=x_2
\right) := F_{n,k,x_1,x_2}(u,v).
\end{align}
On the disjoint event, this is the probability that among $X_3,\dots,X_n$ at least $k$ points fall into $B(x_1,R_n(u))$ and at least $k$ points into $B(x_2,R_n(v))$. Since each point falls into exactly one of the three regions listed above, the count distribution is multinomial with distribution
\begin{align}
F_{n,k,x_1,x_2}(u,v) =&
\sum_{j=k}^{n-2}\sum_{l=k}^{n-2-j}
\binom{n-2}{j}\binom{n-2-j}{l} (P_{n,u,x_1})^j
(P_{n,v,x_2})^l (1-P_{n,u,x_1}-P_{n,v,x_2})^{n-2-j-l}
\end{align}

In the limit, this distribution factorizes as the individual counts converge to Erlang$(k,\bar c\,p(x_i))$ variables. We gloss over the details here and refer the reader to \cite{PS}, only stating the final lemma.

\begin{lemma}[Lemma 47 in \cite{PS}]
    For $x_1 \neq x_2$, we have
    \begin{align}
        \lim_{n \to \infty} F_{n,k,x_1,x_2}(u,v) = F_{k,x_1}(u) F_{k,x_2}(v).
    \end{align}
    with $F_{k,x}$ the CDF of Erlang$(k, \cbar p(x))$, defined as in Lemma~\ref{lem:erlang}.
\end{lemma}

The weak limit $(\zeta(x_1),\zeta(x_2))$ has thus independent components $\zeta(x_i) \sim \mathrm{Erlang}(k,\bar c\,p(x_i))$, and likewise the $q$-sample limit $(\theta(x_1),\theta(x_2))$ has independent components. Moreover, for any $n,m$, the vector $(\zeta_n(x_1),\zeta_n(x_2))$ is a measurable function of only $X^n$ and $(\theta_m(x_1),\theta_m(x_2))$ a measurable function of only $Y^m$. Altogether, the four limit variables $\zeta(x_1),\zeta(x_2),\theta(x_1),\theta(x_2)$ are mutually independent. Consequently,
\begin{align}
&\lim_{n,m\to\infty}\E[W_1 W_2 \mid X_1 = x_1, X_2 = x_2] \\
=&\ \Bigl(\lim_{n\to\infty}\E[W_1 \mid X_1 = x_1]\Bigr)
   \Bigl(\lim_{m\to\infty}\E[W_2 \mid X_2 = x_2]\Bigr).
\end{align}

Since $W_1$ and $W_2$ are identically distributed and convergent, we get (by Lemma~\ref{lem:momconv} and \ref{lem:erlanglog})
\begin{align}
\lim_{n,m\to\infty}\E[W_1 W_2]
=& \left(\lim_{n,m\to\infty}\int_{\cM} \E[W_1 \mid X_1 = x]\,p(x)\,dx\right)^{\!2} = b_s^2,
\end{align}
giving us $\mathbb E[B_n^2]\to b_s^2$. Expanding the rest of the square, we get
\begin{align}
    \lim_{n,m \to \infty} (\E[B_n^2] - 2\E[B_n]b_s + b_s^2) = b_s^2 - 2b_s.b_s + b_s^2 = 0
\end{align}
resulting in $\operatorname{Var}(B_n)\to 0$.
\end{proof}

The next step is to establish convergence in probability of both $A_n$ and $B_n$ to $a_s$ and $b_s$, respectively. This is done by a simple application of Chebyshev's inequality \cite{durrett}.

\begin{lemma}\label{lem:cheb}
    For $k \ge 5$ we have
    \begin{align}
        A_n \xrightarrow{\mathbb P} a_s, \qquad B_n \xrightarrow{\mathbb P} b_s
    \end{align}
\end{lemma}
\begin{proof}
Fix $\varepsilon > 0$. We can decompose $\E[(A_n - a_s)^2]$ with a bias-variance split. Adding and subtracting $\E[A_n]$ inside the square
\begin{align}
    \E\bigl[(A_n - a_s)^2\bigr]
    &= \E\Bigl[\bigl((A_n - \E[A_n]) + (\E[A_n] - a_s)\bigr)^2\Bigr] \\
    &= \E\bigl[(A_n - \E[A_n])^2\bigr]
      + 2(\E[A_n] - a_s)\,\E\bigl[A_n - \E[A_n]\bigr]
      + (\E[A_n] - a_s)^2
\end{align} 
Since $\E[A_n - \E[A_n]] = 0$, the cross-term vanishes and we get $\E[(A_n - a_s)^2] = \Var(A_n) + (\E[A_n] - a_s)^2$. \\

Applying Chebyshev's inequality
\begin{align}
    \mathbb P\bigl(|A_n - a_s| > \varepsilon\bigr)
    &\le \frac{\E[(A_n - a_s)^2]}{\varepsilon^2} \\
    &= \frac{\Var(A_n) + (\E[A_n] - a_s)^2}{\varepsilon^2} .
\end{align}
By Lemmas~\ref{lem:EAn} and~\ref{lem:varAn}, $\E[A_n] \to a_s$ and $\Var(A_n) \to 0$, so the bound vanishes as $n,m \to \infty$ and $A_n \xrightarrow{\mathbb P} a_s$. The argument for $B_n$ is identical, using Lemma~\ref{lem:EBn} and \ref{lem:varBn}, the latter requiring the $k \ge 5$.
\end{proof}

\begin{lemma}\label{lem:cmt}
    If $A_n \xrightarrow{\mathbb P} a_s$ and $B_n \xrightarrow{\mathbb P} b_s$, for $a_s \ne 0$, we have
    \begin{align}
         \frac{B_n}{A_n} \xrightarrow{\mathbb P} \frac{b_s}{a_s}.
    \end{align}
\end{lemma}
\begin{proof}
    Trivially follows from the Continous Mapping theorem \cite{dervaart}.
\end{proof}

Lemmas~\ref{lem:cheb} and \ref{lem:cmt} give us the convergence in probability of the ratio. The only step remaining is to show convergence also in \textit{expectation.} We show this with a uniform integrability argument and applying Vitali's theorem \cite{vitali}.

\begin{lemma}[Uniform integrability of the ratio family]\label{lem:ui}
Assuming $A_n \xrightarrow{\mathbb P} a_s, a_s >0, k \ge 5$ the family
\begin{align}
    \left\{\frac{B_n}{A_n}\right\}_{n\geq 1}
\end{align}
is uniformly integrable, i.e.
\begin{align}
\lim_{M\to\infty}\;\sup_{n,m}\;
\int_{\left\{\left|\frac{B_n}{A_n}\right|>M\right\}}
\left|\frac{B_n}{A_n}\right|\,
dP^{\otimes n}dQ^{\otimes m} = 0 .
\end{align}
\end{lemma}
\begin{proof}
By the de la Vallée-Poussin characterization \cite{poussin}, a family is uniformly integrable whenever a superlinear moment is bounded uniformly; it therefore suffices to show for some $\varepsilon > 0$,
\begin{align}
    \sup_{n,m} \E\!\left[\left|\frac{B_n}{A_n}\right|^{1+\varepsilon}\right] < \infty .
\end{align}
Since $A_n > 0$ almost surely, $|B_n/A_n|^{1+\varepsilon}
= |B_n|^{1+\varepsilon} A_n^{-(1+\varepsilon)}$, and Cauchy--Schwarz \cite{lieb} gives
\begin{align}
    \E\!\left[\left|\frac{B_n}{A_n}\right|^{1+\varepsilon}\right]
    &= \E\!\left[|B_n|^{1+\varepsilon} A_n^{-(1+\varepsilon)}\right] \\
    &\le \E\!\left[|B_n|^{2(1+\varepsilon)}\right]^{1/2}
        \E\!\left[A_n^{-2(1+\varepsilon)}\right]^{1/2} .
\end{align}
We bound the two factors separately, uniformly in $n$. The map $t \mapsto t^{-2(1+\varepsilon)}$ is convex on $(0,\infty)$, so Jensen's inequality \cite{durrett} applied to the average
$A_n = \frac1n\sum_i r_i^\gamma$ yields
\begin{align}
    A_n^{-2(1+\varepsilon)}
    = \left(\frac1n\sum_{i=1}^n r_i^\gamma\right)^{-2(1+\varepsilon)}
    \le \frac1n\sum_{i=1}^n r_i^{-2\gamma(1+\varepsilon)} .
\end{align}
Taking expectations and using that $r_i$ are identically distributed,
\begin{align}
    \E\!\left[A_n^{-2(1+\varepsilon)}\right]
    \le \frac1n\sum_{i=1}^n \E\!\left[r_i^{-2\gamma(1+\varepsilon)}\right]
    = \E\!\left[r_1^{-2\gamma(1+\varepsilon)}\right] ,
\end{align}
which is finite for any $n$ by Lemma~\ref{lem:rmom}, provided $\varepsilon$ is small enough s.t $2\gamma(1+\varepsilon) < k$. Similarly, the map $t \mapsto t^{2(1+\varepsilon)}$ is convex, so Jensen's inequality \cite{durrett} applied to $B_n$ gives
\begin{align}
    |B_n|^{2(1+\varepsilon)}
    &= \left|\frac1n\sum_{i=1}^n r_i^\gamma \log r_i\right|^{2(1+\varepsilon)} \\
    &\le \frac1n\sum_{i=1}^n \left|r_i^\gamma \log r_i\right|^{2(1+\varepsilon)} \\
    &= \frac1n\sum_{i=1}^n r_i^{2\gamma(1+\varepsilon)}
        \left|\log r_i\right|^{2(1+\varepsilon)} .
\end{align}
Taking expectations and using that $r_i$ are identically distributed,
\begin{align}
    \E\!\left[|B_n|^{2(1+\varepsilon)}\right]
    \le \E\!\left[r_1^{2\gamma(1+\varepsilon)}
        \left|\log r_1\right|^{2(1+\varepsilon)}\right] ,
\end{align}
finite and independent of $n$ by Lemma~\ref{lem:logmom}, again for $2\gamma(1+\varepsilon) < k$. For $0<\varepsilon < \frac{k}{2\gamma} - 1$ (non-empty since $k \ge 5$ and $ \gamma < 1)$ both factors are bounded and
\begin{align}
    \sup_{n,m} \E\!\left[\left|\frac{B_n}{A_n}\right|^{1+\varepsilon}\right] < \infty .
\end{align}
\end{proof}
By the previous lemma and Vitali's theorem \cite{vitali} we conclude

\begin{align}
    \lim_{n,m\to\infty}\E\!\left[\left|\frac{B_n}{A_n} - \frac{b_s}{a_s}\right|\right] = 0.
\end{align}

The last remaining step is to plug $b_s/a_s$ into the expectation. Substituting $a_s = \Binv C_s(P,Q)$ and $b_s = \Binv C_s(P,Q)\bigl(G_{k,s} - D_s(P,Q)\bigr)$,
\begin{align}
\frac{b_s}{a_s}
= \frac{\Binv C_s(P,Q)\bigl(G_{k,s} - D_s(P,Q)\bigr)}{\Binv C_s(P,Q)}
= G_{k,s} - D_s(P,Q).
\end{align}
Substituting $b_s/a_s$ 
\begin{align}
    \hat D_s - D_s(P,Q) = \left(G_{k,s} - \frac{B_n}{A_n}\right) - \bigl(G_{k,s} - \tfrac{b_s}{a_s}\bigr) = -\left(\frac{B_n}{A_n} - \frac{b_s}{a_s}\right),
\end{align}
giving us
\begin{align}
    \lim_{n,m\to\infty}\E\bigl[\,\bigl|\hat D_s(X^n,Y^m) - D_s(P,Q)\bigr|\,\bigr] = 0.
\end{align}
Accordingly, $\hat D_s(X^n, Y^m)$ is an $L_1$-consistent estimator for $D_s(P,Q)$.

\subsection{Theorem~\ref{thm:L2}: $L_2$-Consistency}

Recall the $L_2$ consistency statement and simplify the terms inside the expectation:
\begin{align}
    \lim_{n,m\to\infty}\E\!\left[(\hat D_s - D_s)^2\right]
    &= \lim_{n,m\to\infty}\E\!\left[\left(G_{k,s} - \tfrac{B_n}{A_n} - D_s\right)^2\right] \\
    &= \lim_{n,m\to\infty}\E\!\left[\left(\tfrac{B_n}{A_n} - (G_{k,s} - D_s)\right)^2\right] \\
    &= \lim_{n,m\to\infty}\E\!\left[\left(\tfrac{B_n}{A_n} - \tfrac{b_s}{a_s}\right)^2\right]
\end{align}

As a result, if we can prove that the ratio $B_n / A_n \xrightarrow{L_2} b_s /a_s$, we can conclude the $L_2$-consistency. Lemmas~\ref{lem:varAn} and \ref{lem:varBn} (along with \ref{lem:EAn} and \ref{lem:EBn}) give us $A_n \xrightarrow{L_2} a_s$ and $B_n \xrightarrow{L_2} b_s$. To pass to the ratio, we invoke Vitali's theorem again by showing uniform integrability of the squared family.

\begin{lemma}[Uniform integrability of the squared ratio family]\label{lem:uisq}
Assuming $A_n \xrightarrow{\mathbb P} a_s, a_s > 0$ and $ k \ge 5$, the family
\begin{align}
    \left\{\left(\frac{B_n}{A_n}\right)^{2}\right\}_{n\geq 1}
\end{align}
is uniformly integrable.
\end{lemma}
\begin{proof}
The argument is identical to Lemma~\ref{lem:ui}, now applied to the squared family. By de la Vallée-Poussin \cite{poussin} it suffices to bound
$\sup_{n,m} \E[|B_n/A_n|^{2(1+\varepsilon)}]$ for some $\varepsilon>0$, and Cauchy--Schwarz gives
\begin{align}
    \E\!\left[\left|\frac{B_n}{A_n}\right|^{2(1+\varepsilon)}\right]
    \le \E\!\left[|B_n|^{4(1+\varepsilon)}\right]^{1/2}
        \E\!\left[A_n^{-4(1+\varepsilon)}\right]^{1/2} .
\end{align}

Following a similar application of Jensen's inequality, we bound both factors by the $r_i$ moments,
\begin{align}
    &\E\!\left[|B_n|^{4(1+\varepsilon)}\right] \le \E[r_1^{4\gamma(1+\varepsilon)}|\log r_1|^{4(1+\varepsilon)}] \text{ and } \\
    &\E\!\left[A_n^{-4(1+\varepsilon)}\right] \le E[r_1^{-4\gamma(1+\varepsilon)}]
\end{align}
Both are finite by Lemmas~\ref{lem:rmom}
and~\ref{lem:logmom}, provided $4\gamma(1+\varepsilon)<k$. The doubling from the squared family replaces the exponent $2\gamma(1+\varepsilon)$ of Lemma~\ref{lem:ui}
by $4\gamma(1+\varepsilon)$, so the admissible range
$0<\varepsilon<\frac{k}{4\gamma}-1$ is nonempty only for $k \ge 5$.
\end{proof}

Since we already have from \ref{lem:cmt} $B_n / A_n \xrightarrow{\mathbb P} b_s/a_s$, applying Vitali's theorem \cite{vitali} shows $\hat D_s(X^n, Y^m)$ is also an $L_2$ consistent estimator for $D_s(P,Q)$. \qed

\newpage
\section{Additional Experiments}\label{app:exp}
\subsection{Gaussians}
\begin{figure}[H]
    \centering
    \includegraphics[width=\linewidth]{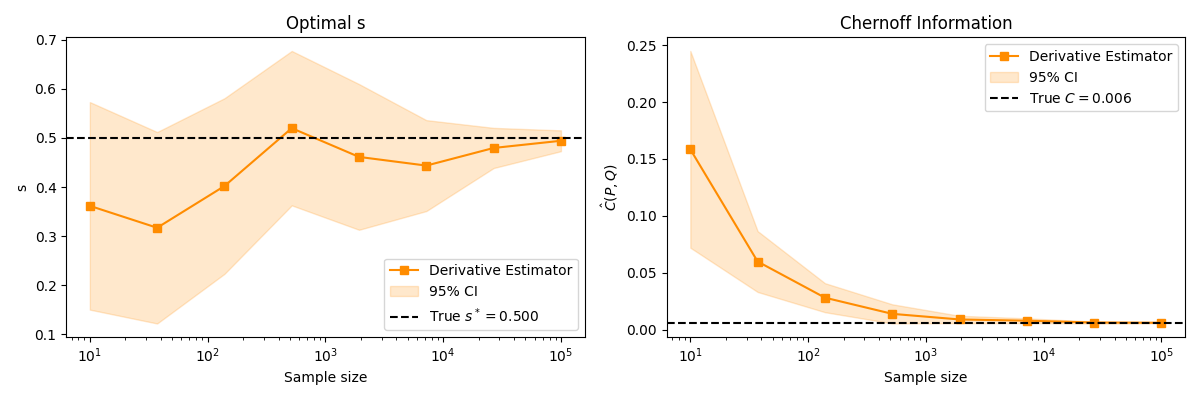}
    \caption{Higher deviation truncated Gaussians $P\sim \cN_{[0,1]}(0.4, 0.5)$ vs. $Q\sim \cN_{[0,1]}(0.6, 0.5)$. As the overlap of the densities is very high, ground truth $C(P,Q)$ is close to $0$. Nevertheless, convergence rate is not effected.}
    \label{fig:largedeviation}
\end{figure}

\begin{figure}
    \centering
    \includegraphics[width=\linewidth]{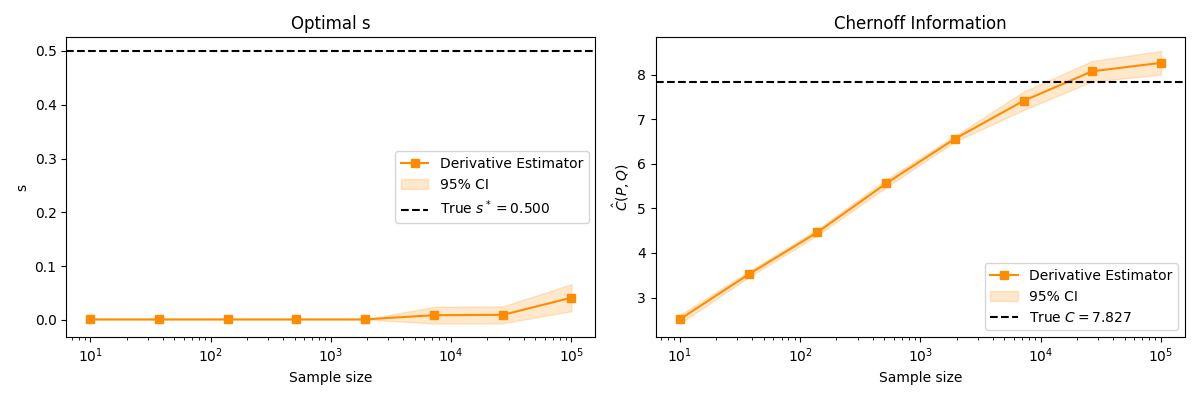}
    \caption{Strongly shifted means on truncated Gaussians, $P\sim \cN_{[0,1]}(0.1, 0.1)$ vs. $Q\sim \cN_{[0,1]}(0.9, 0.1)$ ($\delta = 0.4$). The densities barely overlap, so $\hat D_s$ keeps one sign across $[0,1]$ and the bisection degenerates to $s^* = 0$. Only around $n = 10^5$ does $s^*$ detach from zero suggesting the sample complexity is higher.}
    \label{fig:shiftedmeansnorm}
\end{figure}

\begin{figure}
    \centering
    \includegraphics[width=\linewidth]{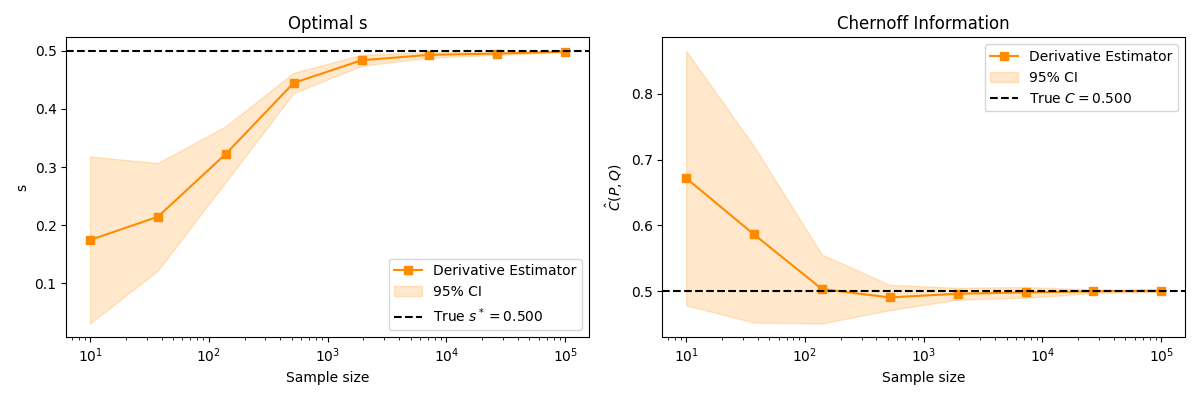}
    \caption{Two regular Gaussians $P\sim \cN(1,1)$ vs $Q \sim \cN(-1,1)$ on unbounded domain $\R$. These distributions violate both Assumption~\ref{ass:bdd} and~\ref{ass:dens}. Interestingly, we observe the same convergence behavior with the truncated Gaussians, suggesting a relaxation of the assumptions is possible as discussed in Remark~\ref{rem:assumptions}.}
    \label{fig:unboundednorm}
\end{figure}

\begin{figure}[H]
    \centering
    \includegraphics[width=\linewidth]{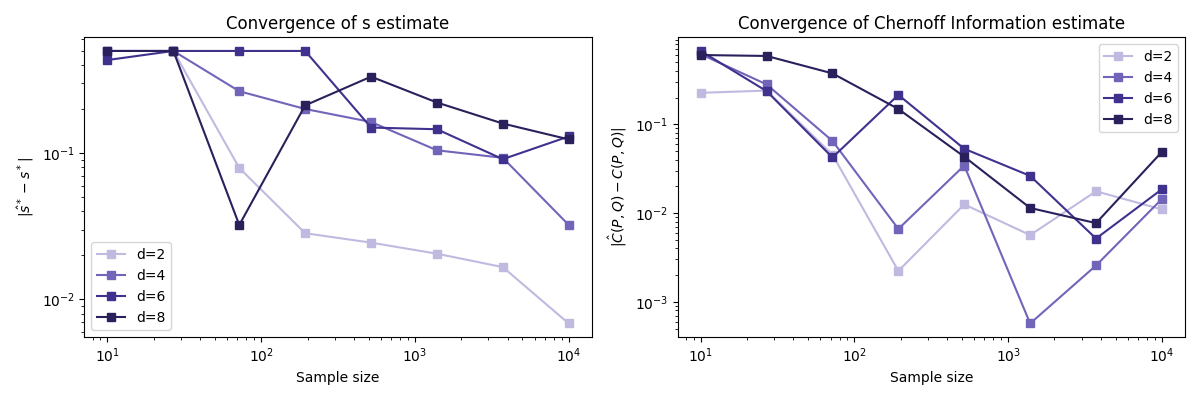}
    \caption{Convergence rate vs dimensionality of truncated Gaussians. Each curve is for a pair of distributions on $\cM^d = [0,1]^d$ with $P \sim \cN_{[0,1]}([0.4]^d, [0.2]^d)$ and $Q \sim \cN_{[0,1]}([0.6]^d, [0.2]^d)$ (symmetric means, $\delta = 0.1$, so $s^* = 1/2$). Here both the $s^*$ and $\widehat C_s$ convergence degrade as dimension increases, in contrast to the exponential case where the $s^*$ converged at similar rates across dimensions. }
    \label{fig:normdim}
\end{figure}

Figures~\ref{fig:largedeviation} --~\ref{fig:normdim} present further experiments with Gaussians: \ref{fig:largedeviation} increases the deviation parameter to cause densities to highly overlap and drive $C(P,Q)$ low, \ref{fig:shiftedmeansnorm} separates them significantly by moving the means apart and \ref{fig:unboundednorm} shows the case when $\cM = \R$, breaking assumptions \ref{ass:bdd} and \ref{ass:dens}. Finally \ref{fig:normdim} studies the higher dimensional case, similar to Figure~\ref{fig:truncexp_dims} from the main section.

\subsection{Exponentials}
\begin{figure}
    \centering
    \includegraphics[width=\linewidth]{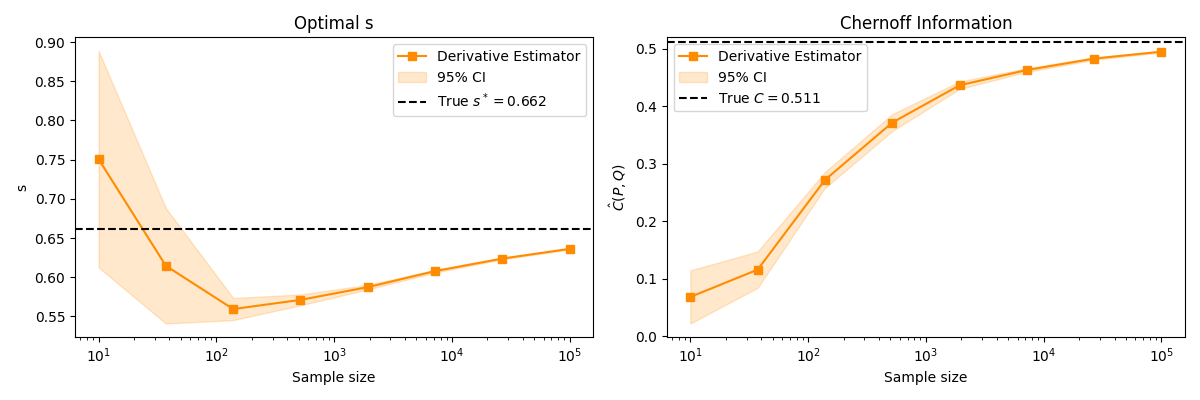}
    \caption{A larger domain, $P \sim Exp_{[0,10]}(1.0)$ vs. $Q \sim Exp_{[0,10]}(8.0)$. At fixed sample size, the wider spread lowers the local sample density, enlarging the $k$-NN volumes and amplifying the finite-sample bias, so estimation quality drops relative to $[0,1]$.}
    \label{fig:largerdom}
\end{figure}

\begin{figure}
    \centering
    \includegraphics[width=\linewidth]{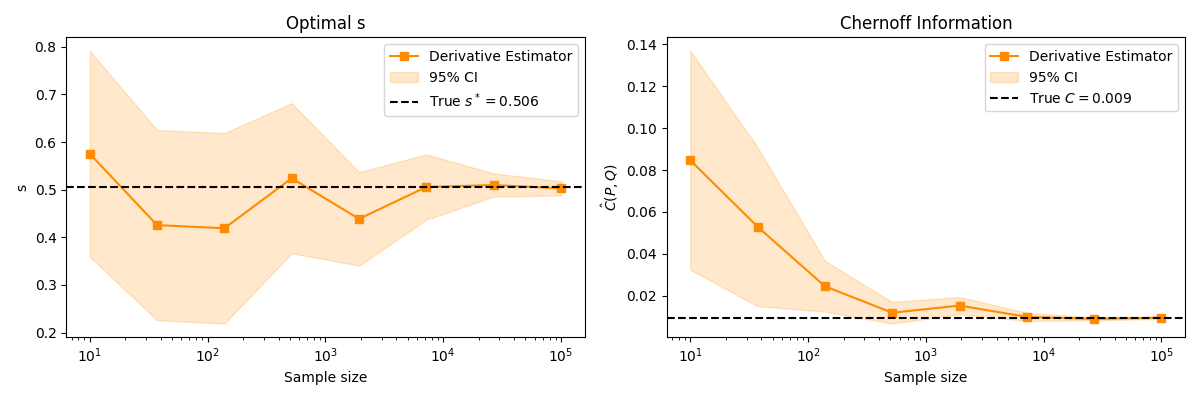}
    \caption{Smaller difference between rates, $P \sim Exp_{[0,1]}(1.0)$ vs. $Q \sim Exp_{[0,1]}(2.0)$. The mixing parameter $s^*$ is very close to $1/2$ and the true Chernoff information is close to $0$. Convergence rate is not significantly effected.}
    \label{fig:smallerrate}
\end{figure}

\begin{figure}[ht]
    \centering
    \includegraphics[width=\linewidth]{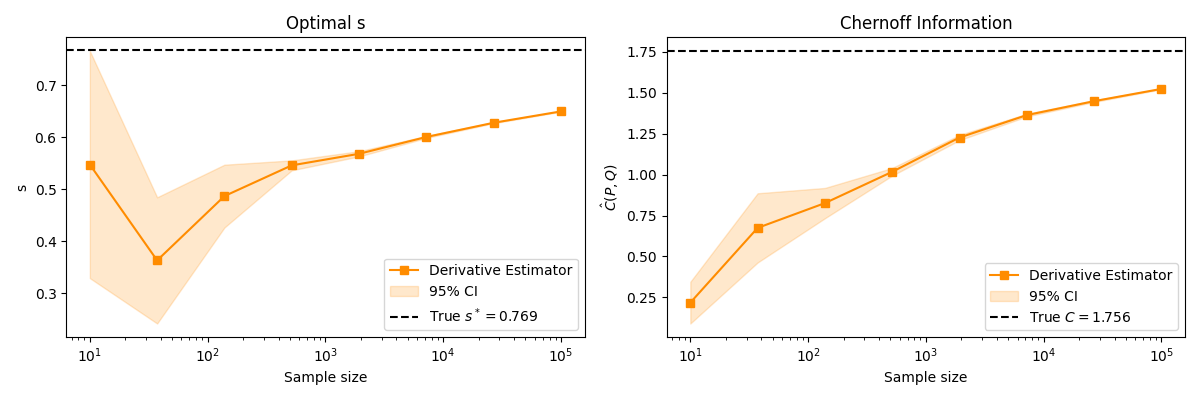}
    \caption{Extreme rate separation, $P \sim Exp_{[0,1]}(1.0)$ vs. $Q \sim Exp_{[0,1]}(100.0)$. Here $q$ concentrates nearly all its mass in $[0, 0.05]$, so $Q$-samples cluster there while $p$ stays spread over $[0,1]$. In the tail where $p$ has mass but $q$ does not, the nearest $Q$-neighbors are far, which degrades the estimation quality.}
    \label{fig:largerate}
\end{figure}

Figures~\ref{fig:largerdom} - \ref{fig:largerate} present the effect of different parameter combinations: \ref{fig:largerdom} increases the domain size while \ref{fig:smallerrate} and \ref{fig:largerate} change the rate parameters $\lambda$ relative to each other.

\end{document}